\documentclass[preprint]{emulateapj} 

\usepackage{natbib}
\usepackage{hyperref}
\hypersetup{colorlinks,citecolor=Blue,linkcolor=Red,urlcolor=Blue}
\usepackage{amsmath}
\usepackage{empheq}
\usepackage{mathrsfs}
\usepackage[usenames,dvipsnames]{color}
\allowdisplaybreaks

\newcommand{\icarus}{Icarus}
\newcommand{\Poincare}{{Poincar$\acute{\rm{e}}$}}
\newcommand{\appropto}{\mathrel{\vcenter{\offinterlineskip\halign{\hfil$##$\cr\propto\cr\noalign{\kern2pt}\sim\cr\noalign{\kern-2pt}}}}}
\begin{document}
 
\title{Dynamical Evolution of Multi-Resonant Systems: the Case of GJ876}  

\author{Konstantin Batygin$^{1}$, Katherine M. Deck$^{1}$, Matthew J. Holman$^{2}$} 
\affil{$^1$Division of Geological and Planetary Sciences, California Institute of Technology, 1200 E. California Blvd., Pasadena, CA 91125}
\affil{$^2$Harvard-Smithsonian Center for Astrophysics, 60 Garden St., Cambridge, MA 02138} 
 
\begin{abstract}

The GJ876 system was among the earliest multi-planetary detections outside of the Solar System, and has long been known to harbor a resonant pair of giant planets. Subsequent characterization of the system revealed the presence of an additional Neptune mass object on an external orbit, locked in a three body Laplace mean motion resonance with the previously known planets. While this system is currently the only known extrasolar example of a Laplace resonance, it differs from the Galilean satellites in that the orbital motion of the planets is known to be chaotic. In this work, we present a simple perturbative model that illuminates the origins of stochasticity inherent to this system and derive analytic estimates of the Lyapunov time as well as the chaotic diffusion coefficient. We then address the formation of the multi-resonant structure within a protoplanetary disk and show that modest turbulent forcing in addition to dissipative effects is required to reproduce the observed chaotic configuration. Accordingly, this work places important constraints on the typical formation environments of planetary systems and informs the attributes of representative orbital architectures that arise from extended disk-driven evolution.

\end{abstract}

\section{Introduction}

The realization that planets form in gaseous protoplanetary disks dates back to the collective works of Swedenborg, Kant and Laplace. Despite several alterations, this nebular hypothesis has survived the test of time \citep{Safronov, Wetherill}. Nevertheless, a quantitative characterization of the consequences of planet-disk interactions \citep{KleyNelson2012} and the associated sculpting of orbital architectures of planetary systems \citep{Morby2013} has only become an active field of research comparatively recently. 

Large-scale orbital migration of giant planets was first recognized as a theoretical possibility by \citet{GoldreichTremaine1980}, in an effort to quantify the orbital evolution of satellites embedded in circum-planetary disks. However, it wasn't until the discovery of the first close-in extrasolar planets \citep{MayorQueloz1995} that this idea gained wide-ranging traction \citep{Lin1996}. Accordingly, over the last two decades, disk-driven migration has been repeatedly invoked as a unifying mechanism to explain the various orbital properties of extrasolar giant planets \citep{LeePeale2002,Crida2007,Batygin2012} as well as low-mass compact multi-planetary systems alike \citep{Papaloizou2007}. 

Qualitatively distinct modes of migration\footnote{Under simplifying assumptions, the two most-commonly quoted categories of migration are that associated with the viscous evolution of the disk (characteristic of giant planets that are able to clear out substantial gaps in their orbital neighborhood) and that facilitated by resonant interactions (characteristic of low-mass planets that remain immersed in the gas).}, characterized by different timescales and directions, can arise depending on the physical properties and structure of the disk as well as the embedded planet \citep{Ward1997,Crida2007,Paardekooper2008,Paardekooper2009,Bert2011,Bert2013}. Because of this intrinsic diversity in physical behavior, simultaneous migration of multiple planets residing within the same disk may cause the orbits to approach each other.

Convergent orbital evolution can result in capture of objects into mean-motion resonances (orbital states characterized by rational period ratios) and evidence for this process is plentiful throughout the satellite population of the Solar System \citep{Goldreich1965,Yoder1973,Yoder1979,Greenberg1977,Peale1976,Peale1986,Henrard1982,HenrardLemaitre1983,Malhotra1990}. Moreover, the existence of a substantial number of (near-)resonant exoplanetary systems (see \citealt{Wright2011,BatyginMorbidelli2013}) suggest that resonant locking is not limited to satellites and is also active among planets \citep{MassetSnellgrove2001,Morby2007}. 

Despite their non-negligible count, resonant systems comprise a minority within the current observational aggregate (\citealt{Wright2011}; see also \citealt{Fabrycky2012}). Instead, giant planets often reside on dynamically excited orbits that are believed to be a result of planet-planet scattering \citep{RasioFord1996,JuricTremaine2008,Raymond2009}. The connection between processes that occur within the nebular epochs of planetary systems and the onset of large-scale orbital instabilities responsible for sculpting the observed semi-major axis - eccentricity distribution are at present poorly understood \citep{Lega2013}. Nevertheless, it is entirely plausible that planet-planet scattering originates within compact systems assembled by disk-driven migration \citep{BeaugeNesvorny2012,Morby2013}.

It is worthwhile to note that the presently favored view of the Solar System's early dynamical history (see \citealt{Morby2008} for a review) is consistent with the picture delineated above. That is, an evolutionary sequence where the giant planets emerged from the Solar nebula in a multi-resonant configuration that subsequently became unstable, causing the planets to scatter onto their current orbits is broadly consistent with the available observations \citep{Tsiganis2005,Morby2007,Levison2008,BatyginBrownFraser2011,NesvornyMorbidelli2012}.

Numerous insights into the physical processes that operate within birth environments of planetary systems could be obtained if the orbital states at the time of nebular dispersion could be inferred. In practice, this is difficult to do for dynamically relaxed systems because planet-planet scattering is a highly stochastic ``forward" process that tends to erase all memory of the system's initial state. In other words, the characterization of the vast majority of the observational sample can yield only limited information about the primordial nature of the associated planetary systems because their dynamical histories have been chaotically eliminated. Even in the Solar System, where the modeling efforts have enjoyed a multitude of observational restrictions, strong degeneracies with respect to the initial state persist \citep{BatyginBrown2010,Nesvorny2011,BatyginBrownBetts2012}.

On the contrary, resonant planetary systems (which have managed to escape the onset of instability\footnote{See \citet{Raymond2008} for an alternative view-point.}) may in fact yield tangible constraints on the environment within which they formed. As such, they constitute exceptionally high value targets for theoretical inquiries. 

Arguably the most exotic resonant exoplanetary system detected to date is GJ876 \citep{Marcy2001,Rivera2005}, and this remarkable collection of planets will be the primary subject of this paper's study. With three objects locked in resonance, GJ876 comprises the only well-characterized extrasolar multi-resonant chain, although additional multi-resonant systems exist within the \textit{Kepler} data set (e.g. Kepler-16, Kepler-79, KOI-730; \citealt{Fabrycky2012}). While this system exhibits some similarity to the Galilean satellites \citep{Peale1976,Peale1986}, it differs from the Io-Europa-Ganemede system in a crucial aspect: the resonant arguments of GJ876 exhibit vigorous, yet bounded (on multi-Gyr timescales) chaos \citep{Rivera2010,Marti2013}.

The architecture of GJ876 is fully consistent with the picture of conventional disk-driven migration \citep{LeePeale2002,Crida2008}. As a result, the dynamical characterization of GJ876 provides a rare window into the description of orbital properties of planetary systems, as they appear when they emerge from their natal disks. With this notion in mind, performing a theoretical analysis of the system's dynamical state with an eye towards obtaining some insights into the nature of the disk from which this system formed is the goal of this work. 

The paper is structured as follows. In section 2, we examine the qualitative features of the system's orbital architecture and setup the basis for theoretical analysis. In section 3, we construct a perturbative model for the resonant dynamics of the system and elucidate the origins of chaotic motion in an analytical fashion. In section 4, we consider the assembly of the multi-resonant chain in a dissipative and turbulent protoplanetary nebula. We conclude and discuss our results in section 5.

\section{The Physical Setup of the Calculation}

The observational saga of the GJ876 system effectively spans the entire active history of exoplanetary science. The initial detection of a $\sim 2 M_{\rm{Jup}}$ giant planet ``b" in a 61 day orbit dates back to the infancy of large-scale dedicated radial velocity surveys \citep{Delfosse1998,Marcy1998}. The discovery of a $\sim 0.7 M_{\rm{Jup}}$ companion ``c" on a 30 day orbit followed shortly thereafter \citep{Marcy2001}, rendering the GJ876 ``c-b" pair the first mean motion resonance to be identified outside the Solar System. Taking advantage of the observational imprint of resonant coupling, \citet{LaughlinChambers2001} and \citet{RiveraLissauer2001} presented independent analyses of the radial velocity data that accounted for planet-planet interactions and were able to break the $\sin(i)$ degeneracy inherent to radial velocity detections, deriving the system's inclination with respect to the line of sight of $i\simeq 50$ deg. Concurrently, the signal of planet ``b" was confirmed astrometrically by \citet{Benedict2002} (see also \citealt{BeanSeifahrt2009}). 

Detections within the system continued, as an additional close-in $\sim 7.5 M_{\oplus}$ planet ``d", residing on a 2 day orbit was announced by \citet{Rivera2005}. Under the assumption of a 3-planet system, \citet{Correia2010} re-analyzed the available data and with extensive modeling derived a mutual inclination between the resonant planets ``b" and ``c" of $\Delta i \simeq 1$ deg. Moreover, the study of \citet{Correia2010} confirmed the existence of planet ``d" and strongly hinted at the eccentric nature of its orbit (see also \citealt{Baluev2011}). 

\begin{table}
\caption{Adopted Orbital Fit of the GJ876 System}
\centering
\begin{tabular}{l l l l l l l }
\hline\hline
$$ & $M$ ($M_{\odot}$) & a (AU) & e & $\mathcal{M}$ (deg) & $\varpi$ (deg)  \\ [0.5ex] 
\hline
$\star$ & $0.334$  & $-$ & $-$ & $-$ & $-$ \\
$d$ & $2.051\times 10^{-5}$  & 0.0208  &  0.207 & 355 & 234   \\
$c$ & $6.820\times 10^{-4}$    & 0.1296 & 0.256  & 294.59 & 48.76  \\
$b$ & $2.173\times 10^{-3}$  & 0.2083 & 0.032  &325.7 & 50.3 \\
$e$ & $4.385\times 10^{-5}$  & 0.3343 & 0.055 &  335 & 239 \\ [1ex]
\hline
\end{tabular}
\label{frequencies}
\end{table}

The latest advancement in the observational characterization of GJ876 arose from the work of \citet{Rivera2010}, who uncovered yet another resonant $\sim 15 M_{\oplus}$ planet ``e," occupying a 124 day orbit. Further dynamical analysis revised the system inclination to $i\simeq 60$ deg and more importantly, showed that the evolution of the multi-resonant configuration undergoes bounded, yet chaotic variations.

\begin{figure*}[t]
\centering
\includegraphics[width=1\textwidth]{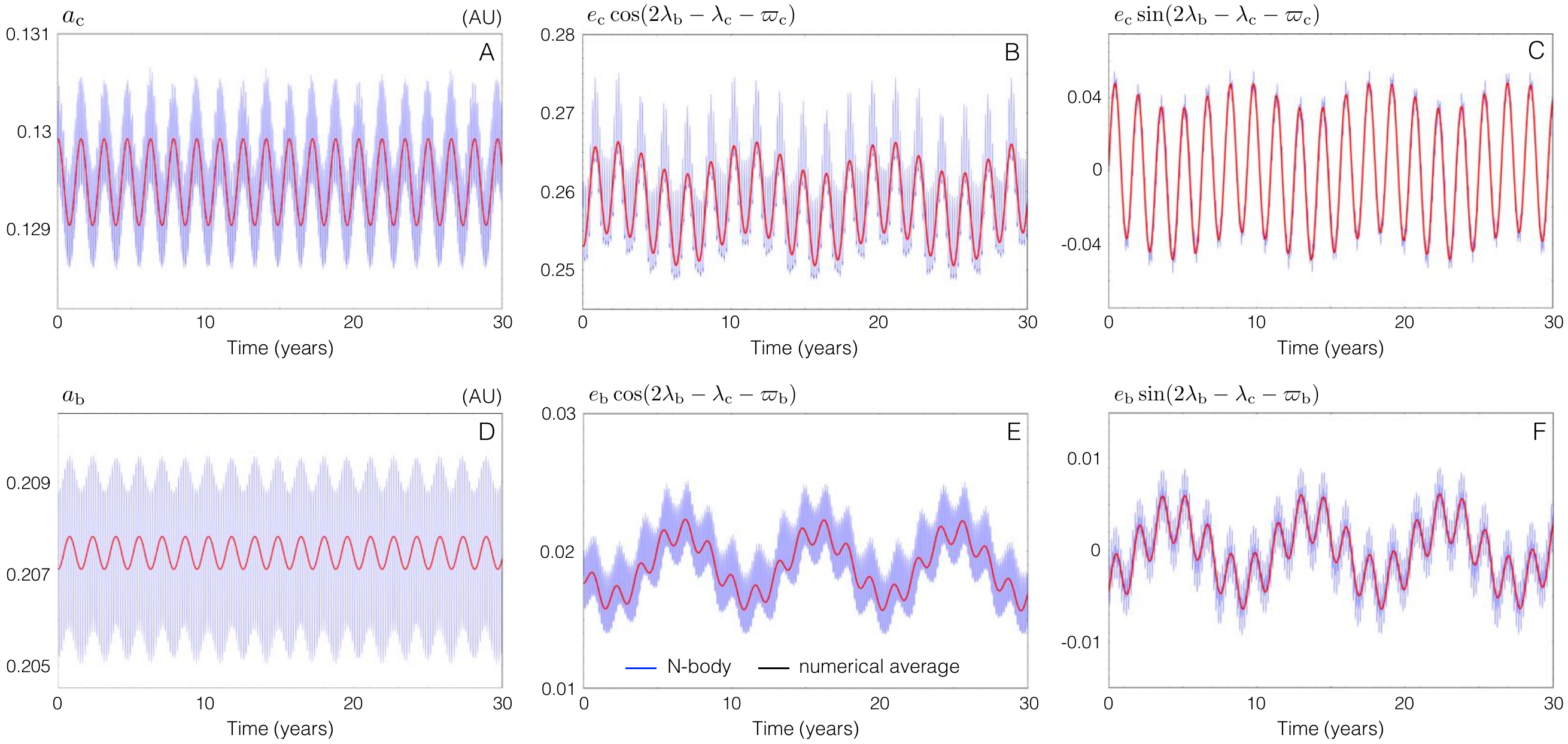}
\caption{Dynamical evolution of the isolated ``c-b" resonant pair. The left panels (A and D) show semi-major axis evolution, while the right panels (B,C,E and F) show eccentricity and critical argument evolution, expressed in terms of scaled canonical cartesian variables. The top and bottom rows corresponds to planets ``c" and ``b" respectively. In each plot, blue curves show osculating elements obtained from direct $N$-body integration. The red curves denote numerically averaged output, where the high-frequency component has been removed by Fourier analysis \citep{Laskar1993,Morby1993}. Treating the outermost planet as a massless test-particle renders the evolution of the massive resonant pair quasi-periodic.}
\label{f1}
\end{figure*}

Over the last decade and a half, the orbital state of the system and its origin have been studied by a substantial number of authors, employing a wide variety of methods. Specifically, in addition to the studies quoted above, orbital characterization and stability have been explored by: \citet{Jones2001,KinoshitaNakai2001,Goz2002,Ji2002,ZhouSun2003,Nader2003,Beauge2003,Veras2007,Baluev2011} and \citet{Marti2013}. Meanwhile the assembly of the particular resonant configuration has been simulated and studied by: \citet{Snellgrove2001,Murray2002,LeePeale2002,ThommesLissauer2003,Beaugeetal2003,Beaugeetal2006, Kley2004,Kley2005,Lee2004,Adams2008,Crida2008} and \citet{LeeThommes2009}. Additionally, \citet{Gerlach2012} examined the possibility of extra bodies being locked in the multi-resonant configuration. 

Thanks to the long time-span covered by the radial velocity observations and the aforementioned fitting efforts, the orbital properties of planets ``b" and ``c" are well constrained \citep{Correia2010}. However, substantial uncertainties exist in the orbital solutions of the lower mass components ``d" and ``e" \citep{Rivera2010}. 

The desired precision of the knowledge of the dynamical state of a system is largely dictated by the purpose of the calculation one wishes to perform. Here, our aim is not to create an ephemeris for GJ876, but rather to shed light on the origins of chaotic motion within the multi-resonant system and place rough constraints on the properties of the nebula within which the system was born. To this end, we note that once a 4-planet system is adopted, chaos (highlighted by the stochastic evolution of planet ``e") is more or less ubiquitous throughout the parameter space restricted by the data \citep{Marti2013}. Consequently, for the purposes of this work, we shall simply adopt the best-fit co-planar orbital solution of \citep{Rivera2010} at face-value, keeping in mind that quantitatively different evolutions stemming from nearby initial conditions can be equally representative of the system's dynamical behavior. To this end, we further remark that while the orbital fit of \citet{Rivera2010} clearly favors a chaotic solution, there may exist quasi-periodic islands in phase space that reside within parameter space covered by observational uncertainties. For completeness, the adopted orbital solution is presented in Table (\ref{f1}). Note that in this work, we have adopted a host stellar mass estimate of \citet{Correia2010}, which differs from that of \citet{Rivera2010} by $\sim 4\%$, meaning that our results will differ from previous works on a detailed level.

\begin{figure*}[t]
\centering
\includegraphics[width=1\textwidth]{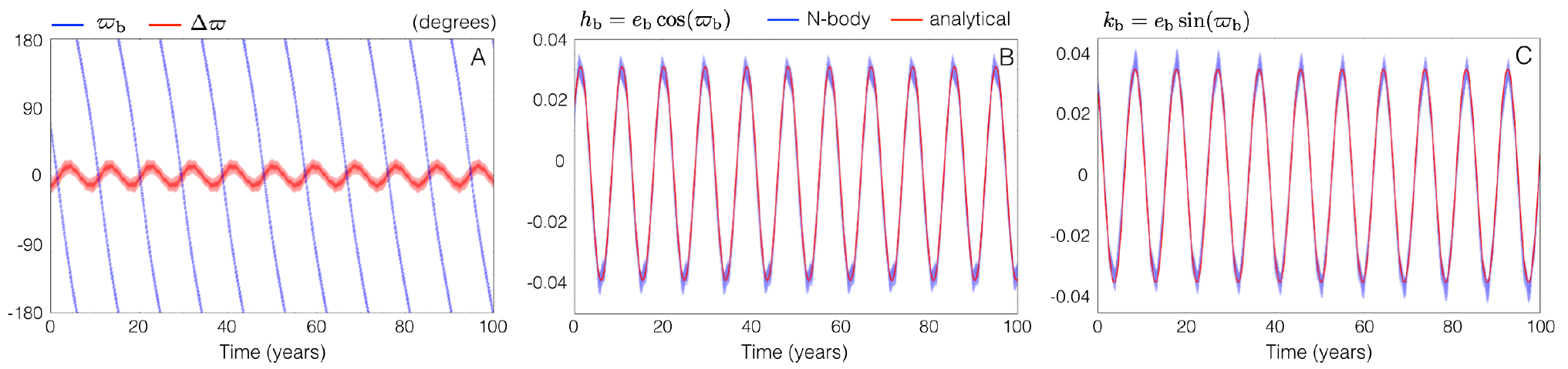}
\caption{Dynamical evolution of planet ``b" in the periodic approximation. The time evolution of the longitude of perihelion of planet ``b" is shown with blue points in panel A. Concurrently, the apsidal difference between planets ``b" and ``c" is shown as a red curve in panel A. Panels B and C depict the evolution of the scaled canonical cartesian coordinates corresponding to planet ``b." As in Figure (1), blue curves represent the time-series obtained from direct numerical integration. Meanwhile, the over plotted red curves show the analytical parameterization given by equation (\ref{hkb}).}
\label{f2}
\end{figure*}

For the purposes of understanding the evolution of the ``c-b-e" multi-resonant chain in the simplest terms possible, the perturbations arising from the close-in planet ``d" can be neglected\footnote{Admittedly, doing so removes some of the high-order features of the system's dynamical portrait. However, detailed scrutiny is unimportant to the problem at hand.}. Broadly speaking, this is justified by the fact that while each member of the chain interacts with its nearest neighbor(s) via first order resonant terms that scale as $\propto e$ (where $e$ is the eccentricity), the averaged gravitational coupling between planet ``d" and the multi-resonant system is secular in nature and to leading order scales as $\propto e^2$ \citep{MD99}. To confirm this argument, we numerically integrated the $N$-body evolution of the system using the \texttt{mercury6} software package \citep{Chambers1999} with and without planet ``d" and found no meaningful differences in the obtained solutions. 

It is further noteworthy that the mass of the outermost planet is more than an order of magnitude smaller than the two inner members of the chain. It is thus tempting to further simplify the system and treat planet ``e" as a test particle, subject to perturbations arising from the ``c-b" resonant pair. Numerical examination of such a setup reveals that qualitative aspects of the dynamical evolution of planet ``e" are largely insensitive to this assumption. On the contrary, under this simplification the evolution of the massive resonant pair ceases to be chaotic and instead exhibits quasi-periodic motion \citep{Beaugeetal2003,Correia2010}. 

Figure (\ref{f1}) shows the results of the simplified integration where planet ``e" is treated as a test particle, and planet ``d" is neglected. Evidently, the stochastic features of the multi-resonant chain arise entirely as a result of the perturbations of planets ``c" and ``b" onto planet ``e," and the chaotic diffusion of the massive pair is communicated exclusively via a back-reaction. This attribute has important consequences for the purposes of constructing a simple model aimed at elucidating the dynamical structure that underlies chaotic motion. Specifically, in this work we will take advantage of this characteristic and derive the chaotic properties of the system by considering the evolution of a massless planet ``e," subject to periodic perturbations from planet ``b," whose orbit is in turn modulated by interactions with planet ``c."

Panel A of Figure (\ref{f2}) shows the numerically obtained time evolution of the longitude of perihelion of planet ``b" (blue) as well as difference between the longitudes of perihelia of planets ``c" and ``b" (red). Clearly, the ``c-b" resonant pair resides near a dynamical equilibrium characterized by alignment and co-precession of the apsidal lines of the orbits \citep{LaughlinChambers2001,LeePeale2002}. Such a state is actually quite peculiar for resonant orbits and is only possible when the orbital eccentricities are not small. Indeed, the classical first-order expansion of the resonant Hamiltonian \citep{LeVerrier1855,EllisMurray2000} does not capture the existence of this fixed point. To this end, \citet{Beaugeetal2003} developed an alternative expansion of the planetary three-body Hamiltonian and showed that the resulting perturbative (first-principles) solution matches the numerically computed evolution well. Taking a somewhat alternative approach, \citet{Correia2010} utilized synthetic perturbation theory to construct a Lagrange-Laplace like periodic solution for the resonant pair which also shows excellent agreement with $N$-body calculations. Because the dynamics of the ``c-b" resonant pair have been studied extensively, here we shall not duplicate the published results and instead refer the interested reader to the aforementioned studies. 

As already briefly mentioned above, in this work we shall examine the consequences of gravitational excitation of planet ``e" facilitated by planet ``b." In order to perform this analysis within a perturbative framework, we must first delineate an approximate functional form for the dynamical behavior of planet ``b." Let us begin by defining the following scaled cartesian canonical coordinates:
\begin{align}
h &= e \cos (\varpi) \nonumber \\
k &= e \sin (\varpi),
\end{align}
where $\varpi$ is the longitude of perihelion. Following \citet{Correia2010}, we note that in the ($h,k$) plane the trajectory of planet ``b" is well described by a circle of radius $\delta$ that is off-set from the origin by $\epsilon$ (see also the discussion on free and forced elements in Ch. 7 of \citealt{MD99}). Accordingly, we parameterize the evolution of planet ``b" in the following manner:
\begin{align}
h_{\rm{b}} &= \epsilon + \delta \cos(g t + \varpi_0) \nonumber \\
k_{\rm{b}} &= \delta \sin(g t + \varpi_0).
\label{hkb}
\end{align}

In the above expression, $g = -0.6706$ rad/yr is the (retrograde) apsidal precession rate and $\varpi_0$ is the phase offset, while the constants are set to $\epsilon = 0.004$ and $\delta = 0.035$. Panels B and C in Figure (\ref{f2}) depict a comparison between results obtained with an $N$-body simulation and the analytical prescription (\ref{hkb}). Needless to say, the observed agreement is excellent. 

It is further noteworthy that the numerically averaged semi-major axis of planet ``b" does not deviate from its nominal value, $[a]_{\rm{b}}$ by more than a few parts in a thousand. Thus, for the purposes of the following calculation we may readily neglect the semi-major axis evolution all together and set $a_{\rm{b}}= [a]_{\rm{b}}$. With a simple analytical model for the dynamical evolution of the perturbing planet at hand, we are now in a position to perform a perturbative analysis of the system's chaotic behavior.

\section{Origins of Chaotic Motion}

\begin{figure*}[t]
\centering
\includegraphics[width=1\textwidth]{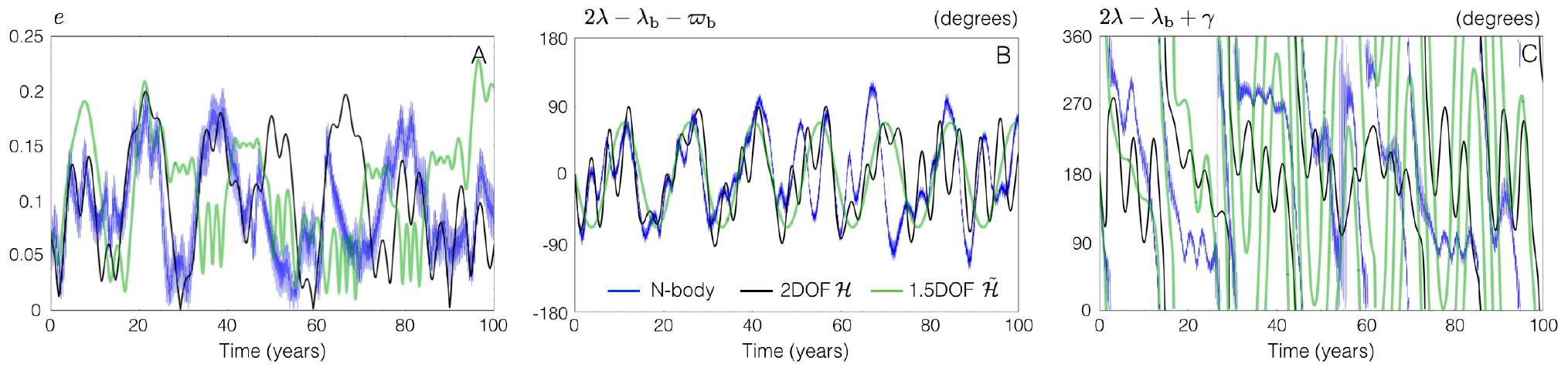}
\caption{Chaotic evolution of the outermost planet, ``e." Panel A shows the eccentricity as a function of time, while panels B and C show the evolution of the critical arguments corresponding to the first-order (2:1) mean motion resonance. Across all panels, the blue curves represent time series obtained by direct $N$-body integration, the black curves denote solutions arising from the autonomous two degree of freedom perturbative model (Hamiltonian \ref{HAa}), and the green curves show the evolution described by the simplified non-autonomous one degree of freedom model (Hamiltonian \ref{HoneDOF}). Note that all solutions track each-other well initially but lose coherence after a few decades of evolution.}
\label{f3}
\end{figure*}

Before we proceed, let us begin by defining restricted \Poincare\ action-angle variables in terms of standard orbital elements: 
\begin{align}
\Lambda &= \sqrt{\mathcal{G} M_{\star} a} \ \ \ \ \ \ \ \ \ \ \ \ \ \ \ \ \ \ \ \ \ \ \ \ \ \ \lambda = \mathcal{M} + \varpi \nonumber \\
\Gamma &= \Lambda (1 - \sqrt{1-e^2}) \simeq [\Lambda] e^2/2 \ \ \ \ \ \ \ \ \gamma = -\varpi
\end{align}
where $\mathcal{G}$ is the gravitational constant, $M_{\star}$ is the stellar mass, $\mathcal{M}$ is the mean anomaly and the nominal action $[\Lambda]$ is evaluated at $[a] = 2^{2/3} [a]_{\rm{b}}$. For the remainder of the paper, quantities not labeled with an index shall correspond to planet ``e." 

\subsection{Resonant Perturbation Theory}

Generally, unlike secular perturbations (see e.g. \citealt{Laskar1996}), mean-motion resonances modulate both the eccentricity and semi-major axis. Hence, Keplerian motion cannot be averaged over, and the corresponding part of the Hamiltonian must be retained:
\begin{align}
\mathcal{H}_{\rm{kep}} &= - \frac{\mathcal{G}^2 M_{\star}^2}{2 \Lambda^2}.
\label{Hkep}
\end{align}
Expanding $\mathcal{H}_{\rm{kep}}$ around the nominal 2:1 resonant semi-major axis quoted above, equation (\ref{Hkep}) takes the form:
\begin{align}
\mathcal{H}_{\rm{kep}} &= - \frac{\mathcal{G}^2 M_{\star}^2}{2 [\Lambda]^2} + \frac{\mathcal{G}^2 M_{\star}^2}{[\Lambda]^3} \big(\Lambda - [\Lambda]\big) -  \frac{3\mathcal{G}^2 M_{\star}^2}{2[\Lambda]^4} \big(\Lambda - [\Lambda]\big)^2 \nonumber \\
&= - \frac{3 \mathcal{G}^2 M_{\star}^2}{[\Lambda]^2} + 4 [n] \Lambda - \frac{3}{2} [h] \Lambda^2,
\label{Hkep2}
\end{align}
where $[n]=\sqrt{\mathcal{G} M_{\star}/[a]^3}$ is the mean motion and $[h]=[n]/[\Lambda]$. The first term on the second line of equation (\ref{Hkep2}) is constant and can thus be dropped from the Hamiltonian. 

Upon averaging out short-periodic terms (see Ch. 2 of \citealt{MorbyBook}), the component of the Hamiltonian that governs first order (in $e$) resonant planet-planet interactions (within the framework of the elliptic restricted three body problem) reads \citep{MD99}:
\begin{align}
\mathcal{H}_{\rm{res}} &= - \frac{\mathcal{G} M_{\rm{b}}}{[a]} \bigg( f_{\rm{res}}^{(1)} e_{\rm{b}} \cos(2\lambda - \lambda_{\rm{b}} - \varpi_{\rm{b}}) \nonumber \\
&+ f_{\rm{res}}^{(2)} \sqrt{\frac{2 \Gamma}{[\Lambda]}} \cos(2 \lambda - \lambda_{\rm{b}} + \gamma) \bigg),
\label{Hres}
\end{align}
where $f_{\rm{res}}^{(1)} = -1.1905$ and $f_{\rm{res}}^{(2)} = 0.4284$ are interaction coefficients that depend exclusively on the semi-major axis ratio \citep{MD99}. The presence of $e_{\rm{b}}$ and $\varpi_{\rm{b}}$ in the Hamiltonian (\ref{Hres}) gives rise to explicit time-dependence in expression (\ref{Hres}). Thus, the quoted model constitutes a non-autonomous dynamical system with two degrees of freedom. 

The Hamiltonian can be made autonomous by extending the phase-space to three degrees of freedom \citep{MorbyBook}:
\begin{align}
\mathcal{H} &= \mathcal{H}_{\rm{kep}} + \mathcal{H}_{\rm{res}} + \mathbb{T},
\label{HT1}
\end{align}
where $\mathbb{T}$ is the newly introduced action conjugate to time, $t$. Accordingly, substituting the solution (\ref{hkb}) into equation (\ref{Hres}), we obtain the following expression: 
\begin{align}
\mathcal{H} &= 4 [n] \Lambda - \frac{3}{2} [h] \Lambda^2 + \mathbb{T} - \frac{\mathcal{G} M_{\rm{b}}}{[a]} \nonumber \\
&\times \bigg(  f_{\rm{res}}^{(1)} \big( (\epsilon + \delta \cos(g t + \varphi_0)) \cos(2 \lambda - n_{\rm{b}} t - \lambda_{\rm{b}0}) \nonumber \\
&+ \delta \sin(gt - \varpi_0) \sin(2 \lambda - n_{\rm{b}} t - \lambda_{\rm{b}0} ) \big) \nonumber \\
&+ f_{\rm{res}}^{(2)} \sqrt{\frac{2 \Gamma}{[\Lambda]}} \cos(2 \lambda - n_{\rm{b}} t - \lambda_{\rm{b}0} +\gamma) \bigg),
\label{HT}
\end{align}
where $\lambda_{\rm{b}0}$ is an initial phase of planet ``b," and $n_{\rm{b}}$ is its mean motion.

Equation (\ref{HT}) is rather cumbersome and can be made more succinct. First, let us define the constants:
\begin{align}
\alpha &= - \frac{\mathcal{G} M_{\rm{b}}}{[a]} f_{\rm{res}}^{(1)} \nonumber \\
\beta &= - \frac{\mathcal{G} M_{\rm{b}}}{[a]} \frac{f_{\rm{res}}^{(2)}}{\sqrt{[\Lambda]}}.
\label{resconst}
\end{align}
Next, let us perform a canonical transformation of coordinates, arising from the following type-2 generating function \citep{Goldstein}: 
\begin{align}
\mathcal{F}_2 = (2 \lambda - n_{\rm{b}} t - \lambda_{\rm{b}0}) \Theta + (\gamma) \Phi + (t) \Xi.
\label{transone}
\end{align}
An application of the transformation equations yields the new action-angle variables:
\begin{align}
\Theta &= \Lambda/2 \ \ \ \ \ \ \ \ \ \ \ \ \ \ \ \ \ \ \  \theta = 2 \lambda - n_{\rm{b}} t - \lambda_{\rm{b}0} \nonumber \\
\Phi &= \Gamma  \ \ \ \ \ \ \ \ \ \ \ \ \ \ \ \ \ \ \ \ \ \  \phi = \gamma \nonumber \\
\Xi &= \mathbb{T} + n_{\rm{b}} \Theta \ \ \ \ \ \ \ \ \ \ \ \ \ \xi = t.
\end{align}
In terms of the new coordinates, the Hamiltonian reads:
\begin{align}
\mathcal{H} &= 8 [n] \Theta - 6 [h] \Theta^2 + \alpha \epsilon \cos(\theta) + \alpha \delta \cos(\theta) \nonumber \\
&\times \cos(g \xi + \varphi_0) + \alpha \delta \sin(\theta) \sin(g \xi + \varphi_0) \nonumber \\
&+ \beta \sqrt{2 \Phi} \cos(\theta + \phi) - n_{\rm{b}} \Theta + \Xi.
\label{HXI}
\end{align}

Although somewhat less unwieldily than equation (\ref{HT}), the Hamiltonian (\ref{HXI}) is still characterized by three degrees of freedom, precluding straightforward analytical treatment. To remedy this issue, let us define canonical cartesian coordinates related to the $(\Phi,\phi)$ degree of freedom:
\begin{align}
x = \sqrt{2 \Phi} \cos(\phi) \ \ \ \ \ \ \ y = \sqrt{2 \Phi} \sin(\phi).
\label{xy}
\end{align}
After some algebraic manipulation, the Hamiltonian obtains the form: 
\begin{align}
\mathcal{H} &=(8 [n]  - n_{\rm{b}} )\Theta - 6 [h] \Theta^2 + \alpha \delta \cos(\theta - g \xi - \varphi_0) \nonumber \\
&+ \alpha \epsilon \cos(\theta) + \beta x \cos(\theta) - \beta y \sin(\theta) + \Xi.
\label{Hxy1}
\end{align}
To reduce the number of harmonics present in $\mathcal{H}$, we follow \citet{Henrard1986,Wisdom1986} and define the following canonical translation\footnote{Note that for the non-restricted (i.e. planetary) three-body problem, there exists a corresponding canonical rotation that reduces the first-order resonant Hamiltonian to an integrable one \citep{SessinFerrazMello1984,Wisdom1986,BatyginMorbidelli2013,Deck2013,Delisle2014}.}: 
\begin{align}
\tilde{x} = x + \frac{\alpha}{\beta} \epsilon \ \ \ \ \ \ \ \tilde{y} = y.
\label{reducing}
\end{align}
This transformation morphs the first and second terms on the second line of equation (\ref{Hxy1}) into a single term. Accordingly, upon defining implicit action-angle variables
\begin{align}
\tilde{x} = \sqrt{2 \Psi} \cos(\psi) \ \ \ \ \ \ \ \tilde{y} = \sqrt{2 \Psi} \sin(\psi),
\label{psi}
\end{align}
the Hamiltonian takes on a form characterized by only two harmonics:
\begin{align}
\mathcal{H} &=(8 [n]  - n_{\rm{b}} )\Theta - 6 [h] \Theta^2 + \alpha \delta \cos(\theta - g \xi - \varphi_0) \nonumber \\
&+ \beta \sqrt{2 \Psi} \cos(\theta + \psi) + \Xi.
\label{Hxy}
\end{align}

\begin{figure*}[t]
\centering
\includegraphics[width=0.8\textwidth]{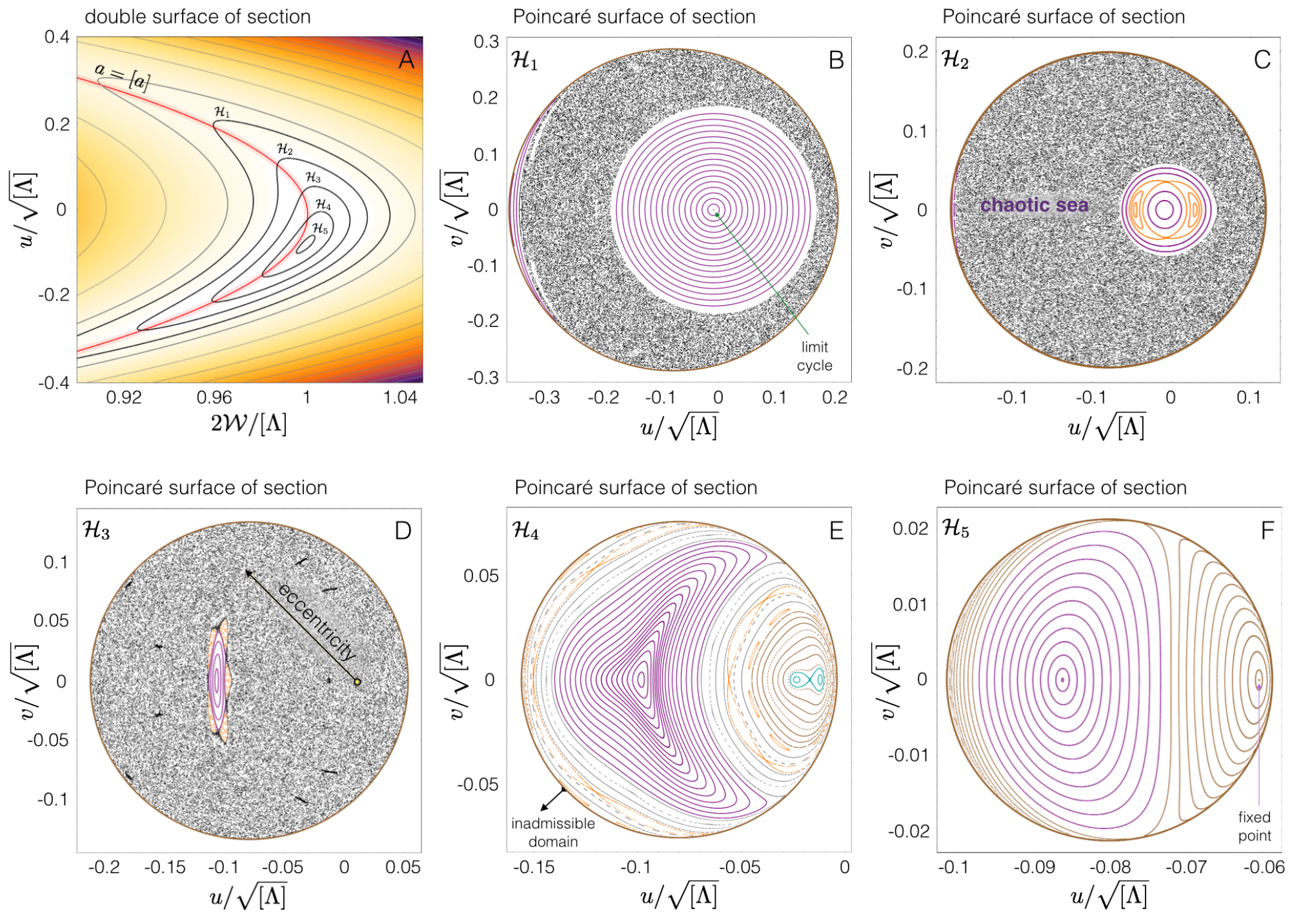}
\caption{Surfaces of section corresponding to the autonomous two degree of freedom Hamiltonian (\ref{HAa}). Panel A shows level curves of the Hamiltonian on a double section, and the rest of the panels depict \Poincare\ surfaces of section with respect to the $(\mathcal{W},w)$ degree of freedom, at various levels of $\mathcal{H}$. The nominal resonant semi-major axis is shown with a red curve in panel A. In the other panels, the variables are scaled such that the radial distance away from the origin approximately corresponds to the eccentricity. On the \Poincare\ surfaces of section, the chaotic sea is shown with black points, while quasi-periodic trajectories are shown with purple, brown, and orange points. Note that chaotic excisions are limited by the conservation of $\mathcal{H}$, which give rise to inadmissible regions in phase-space.}
\label{f4}
\end{figure*}

From here, the reduction of the Hamiltonian to a two-degree of freedom system is straightforward. Particularly as in equation (\ref{transone}), define a contact transformation arising from a type-2 generating function:
\begin{align}
\mathcal{F}_2 = (\theta - g \xi - \varphi_0) \mathcal{W} + (\theta + \psi) \mathcal{Z} + (\xi) \mathcal{K}.
\label{transtwo}
\end{align}
The transformation equations yield:
\begin{align}
\mathcal{W} &= \Theta - \mathcal{Z} \ \ \ \ \ \ \ \ \ \ \ \ \ \ \ \ w = \theta - g \xi - \varphi_0 \nonumber \\
\mathcal{Z} &= \Psi  \ \ \ \ \ \ \ \ \ \ \ \ \ \ \ \ \ \ \ \ \ \  z = \theta + \psi \nonumber \\
\mathcal{K} &= \Xi + g \mathcal{W} \ \ \ \ \ \ \ \ \ \ \ \ \ \ \kappa = \xi.
\label{finangles}
\end{align}
Noting that the resonant condition implies that $n_{\rm{b}} = 2[n]$ the Hamiltonian takes the form: 
\begin{align}
\mathcal{H} &=6 [n] (\mathcal{W} + \mathcal{Z}) - 6 [h] (\mathcal{W} + \mathcal{Z})^2 - g \mathcal{W}  \nonumber \\
&+ \alpha \delta \cos(w) + \beta \sqrt{2 \mathcal{Z}} \cos(z) + \mathcal{K}.
\label{HAaone}
\end{align}
Because the angle $\kappa$ (which corresponds to time) is now absent from the Hamiltonian, the action $\mathcal{K}$ is a constant of motion and can thus be dropped from equation (\ref{HAaone}). 

The reduction of the Hamiltonian to an autonomous two-degree of freedom system is now complete. Qualitatiely, the Hamiltonian (\ref{HAaone}) represents a system of two momentum-coupled pendulums. It is noteworthy that in isolation, the $(\mathcal{Z},z)$ degree of freedom possesses the D'Alembert characteristic and thus has the dynamics corresponding to the second fundamental model for resonance \citep{HenrardLemaitre1983} whereas the isolated $(\mathcal{W},w)$ degree of freedom has the phase-space structure of a simple pendulum \citep{Chirikov1979}.

The extent to which the perturbative model governed by Hamiltonian (\ref{HAaone}) successfully captures the dynamical behavior of GJ876 ``e" can be tested numerically. However prior to applying Hamilton's equations, note that in terms of action-angle coordinates, Hamiltonian (\ref{HAaone}) possesses a coordinate singularity at $\mathcal{Z} = 0$. This nuisance can be circumvented by transforming to canonical cartesian coordinates: 
\begin{align}
u = \sqrt{2 \mathcal{Z}} \cos(z) \ \ \ \ \ \ \ v = \sqrt{2 \mathcal{Z}} \sin(z).
\label{uv}
\end{align}
Accordingly, the Hamiltonian is rewritten as follows:
\begin{align}
\mathcal{H} &=6 [n] \left(\mathcal{W} + \frac{u^2 + v^2}{2} \right) - 6 [h] \left(\mathcal{W} + \frac{u^2 + v^2}{2} \right)^2  \nonumber \\
&- g \mathcal{W} + \alpha \delta \cos(w) + \beta u.
\label{HAa}
\end{align}
The corresponding equations of motion take the form:
\begin{align}
\frac{d\mathcal{W}}{dt} &= \alpha \delta \sin(w) \nonumber \\
\frac{dw}{dt} &= 6 [n] - g - 12 [h] \left( \mathcal{W} + \frac{u^2 + v^2}{2} \right) \nonumber \\
\frac{du}{dt} &= - 6 [n] v + 12 [h] v \left( \mathcal{W} + \frac{u^2 + v^2}{2} \right) \nonumber \\
\frac{dv}{dt} &= \beta + 6 [n] u - 12 [h] u \left( \mathcal{W} + \frac{u^2 + v^2}{2} \right) 
\label{eom}
\end{align}

After numerous variable changes, it is useful to explicitly relate the final variables (\ref{finangles},\ref{uv}) to Keplerian orbital elements. Working back through the transformations delineated above, it can be shown that the orbital semi-major axis and eccentricity are expressed as follows:
\begin{align}
a &= \frac{( 2\mathcal{W} + \left( u - \epsilon(\alpha/\beta) \right)^2 + v^2 )^2}{\mathcal{G} M_{\star} } \nonumber \\
e &= \sqrt{\frac{\left( u - \epsilon(\alpha/\beta) \right)^2 + v^2 }{\sqrt{\mathcal{G} M_{\star} [a]}}}.
\label{var2kep}
\end{align}
Meanwhile, the angles present in the original formulation of the Hamiltonian (\ref{Hres}) are related to the variables (\ref{finangles}) in a unembellished way:
\begin{align}
w &= (2 \lambda - \lambda_{\rm{b}} - \varpi_{\rm{b}}) \nonumber \\
z &= \arctan\left(\frac{v}{u} \right) \simeq (2 \lambda - \lambda_{\rm{b}} + \gamma).
\end{align}
The latter equality is inexact, but is nevertheless a good approximation because $\epsilon \ll \delta$. In the same spirit, the following relationships approximately hold:
\begin{align}
&u \simeq e \cos(z) \nonumber \\
&v \simeq  e \sin(z) \nonumber \\
&\frac{2\mathcal{W}}{[\Lambda]} \simeq \sqrt{\frac{a}{[a]}} - e^2.
\label{var2kepref}
\end{align}
Note that the last of the above expressions is closely related to the well-known Tisserand parameter.

Orbital evolution obtained by numerical integration of equations (\ref{eom}) is shown with black curves in Figure (\ref{f3}), where the results of an $N$-body simulation are also depicted with blue lines. Evidently, the system described by Hamiltonian (\ref{HAa}) provides an excellent perturbative representation of the real system, meaning that the dynamical behavior within the chaotic layer is well captured by a first-order expansion of the disturbing function. It is noteworthy that the semi-analytical solution initially tracks the $N$-body solution, but the two evolutionary sequences lose coherence after $\sim 50$ years. This is indicative of a Lyapunov time that is a factor of a few shorter than $50$ years\footnote{This is in some contrast with a $10^2-10^3$ year Lyapunov time reported by \citet{Rivera2010}.}. This places GJ876 into the same category of rapidly chaotic systems as \textit{Kepler}-36 \citep{Deck2012}. 

Despite rapid chaotic diffusion, $N$-body calculations reported by \citet{Rivera2010} suggest that the system is stable on multi-Gyr timescales. In other words, the chaos exhibited by the multi-resonant system is simultaneously vigorous and bounded. Both of these characteristics can be qualitatively understood within the context of Hamiltonian (\ref{HAa}) by examination of surfaces of section. 

As a starting point, let us examine the evolution of the angles depicted in Figure (\ref{f3}). Note that the angle $(2 \lambda_{\rm{b}} - \lambda - \varpi_{\rm{b}})$ librates around $0$ while $(2 \lambda_{\rm{b}} - \lambda + \gamma)$ switches between libration around $\pi$ and circulation. Preliminary intuition about the amplitude of orbital excursions can be obtained by constructing a double surface of section of the Hamiltonian. In accord with the evolution of angles denoted in Figure (\ref{f3}), we fix $w = 0$ and $v = 0$ and plot level curves of $\mathcal{H}$ on panel A of Figure (\ref{f4}). Suitably, on the $x$-axis, we plot the scaled action $2\mathcal{W}/[\Lambda]$ while on the $y$-axis we plot the scaled momentum $u/\sqrt{[\Lambda]}$. Because the Hamiltonian (\ref{HAa}) is autonomous, any (potentially chaotic) trajectory it describes will be constrained to map onto a corresponding level curve (given by the value of $\mathcal{H}$) every time the orbital state crosses the section condition.

A red curve corresponding to the nominal semi-major axis $a = [a]$ is also shown on this panel. Crudely speaking, the horizontal deviation away from the nominal resonance curve on the double section is indicative of changes in semi-major axis whereas vertical deviation corresponds to eccentricity modulation. Thus, the area occupied by a given constant $\mathcal{H}$ curve on the double section serves as a proxy for the amplitude of orbital variations associated with resonant motion. This point is of some importance to understanding the long-term stability of the system. Particularly, this simple analysis suggests that to the extent that Hamiltonian (\ref{HAa}) provides an adequate representation of the dynamics, the conservation of $\mathcal{H}$ restricts the maximal deviation from nominal resonance of the trajectory, no matter how vigorously chaotic it may be. Qualitatively, this explains how a rapidly stochastic system such as GJ876 can remain stable on multi-Gyr timescales. 

\begin{figure*}[t]
\centering
\includegraphics[width=1\textwidth]{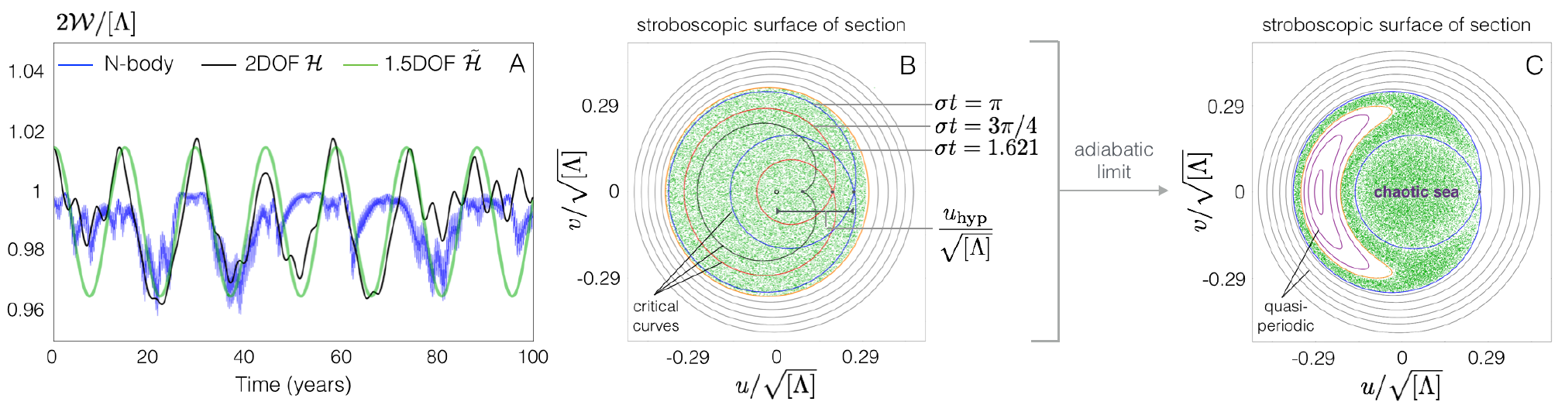}
\caption{Properties of the simplified model governed by Hamiltonian (\ref{HoneDOF}). Panel A depicts the evolution of the action $\mathcal{W}$, where the color scheme is the same as that employed in Figure (3). Note that the evolution corresponding to the non-autonomous system is simply that dictated by equation (\ref{Wt}). Panel B shows a stroboscopic surface of section with the same section conditions as that employed in Figure (4). The sweeping of the separatrix is shown by depiction of its sequential shape at different times. The maximal extent of the hyperbolic equilibrium point ($u_{\rm{hyp}}$) is also explicitly labeled. In order to highlight the non-adiabatic nature of the system, panel C shows the adiabatic limit (where the modulation frequency $\sigma$ is reduced by a factor of 10) of the equation of motion arising from Hamiltonian (\ref{HoneDOF}). Note that reducing $\sigma$ significantly introduces quasi-periodic resonant trajectories.}
\label{f5}
\end{figure*}

In addition to delineating energy contours on a double section, it is further useful to explore the actual dynamical behavior of the system on the contours (i.e. provided specific values of $\mathcal{H}$). To this end, we have constructed \Poincare\ surfaces of section for the five highlighted curves shown on the double section, labeled $\mathcal{H}_1 ... \mathcal{H}_5$. The sections are taken with respect to $w = 0$ at $d w/dt <0$ and depict the $(u,v)$ degree of freedom. The surfaces are shown as panels B-F on Figure (\ref{f4}) and are labeled according to the level of $\mathcal{H}$ they represent. The value of $\mathcal{H}$ that corresponds to the initial condition depicted in Table (\ref{f1}) is $\mathcal{H}_3$ and is shown on panel D.

Quasi-periodic trajectories, shown as purple and brown curves dominate the surfaces of section that correspond to energy levels that yield the most limited orbital variations (i.e. $\mathcal{H}_4$ and $\mathcal{H}_5$, shown on panels E and F respectively). On the contrary, the remaining surfaces of section feature substantial chaotic seas, shown with black points. The location of a given point on the surface of section is related to the orbital eccentricity (evaluated at the section conditions) through expression (\ref{var2kep}). Accordingly, the range of chaotic eccentricity excursions at a given energy level can be gathered from examining the extent of the phase-space occupied by the irregular region. Note further that in all cases, the dynamics resides on a well-defined domain, which is restricted by the conservation of $\mathcal{H}$ and the requirement for the actions to have a null imaginary component. 

The physical characters of the periodic points that reside at the centers of the nested quasi-periodic trajectories is not uniform across the plotted energy levels. Particularly, on panels B and C, these orbits correspond to limit cycles that are characterized by rapid circulation of $w$ and a modulated circulation of $z$, such that the majority of time spent on the circulation cycle remains in the vicinity of $z=\pi$. Conversely, fixed points in panels D, E, and F are true equilibria that are characterized by stationary evolution of the angles at $w=0$ and $z=\pi$.

Despite these differences, as long as the system is initiated within the extensive chaotic region (irrespective of whether the value of the Hamiltonian is set to $\mathcal{H}_1$, $\mathcal{H}_2$, or $\mathcal{H}_3$), the qualitative behavior of the dynamics resembles that observed in Figure (\ref{f3}). In light of this, it is tempting to obtain an estimate for the characteristic rate of chaotic decoherence of the system, that is independent of the exact value of $\mathcal{H}$. The simplest approach to such an estimate requires further reduction of the model.

\subsection{A Simplified Model}

Recall that the only assumptions inherent to the Hamiltonian (\ref{HAa}) are the truncation of the disturbing function at first order in $e$ \citep{MD99} and the adoption of equations (\ref{hkb}) for the dynamical evolution of the perturbing planet. The subsequent conversion of the Hamiltonian to an autonomous two degree of freedom system was possible because of the reducing transformation (\ref{reducing}) \citep{Henrard1986,Wisdom1986}. Because the Hamiltonian (\ref{HAa}) cannot be simplified further with canonical transformations, in order to convert the system into a more tractable non-autonomous one degree of freedom system (see e.g. \citealt{Timofeev1978,Escande1985}), we must prescribe a functional form to one of the degrees of freedom.

An examination of panels B and C of Figure (\ref{f3}) shows that while the oscillations of the angle $w$ are approximately limited to the range $-\pi/2 \lesssim w \lesssim \pi/2$, the angle $z$ recurrently transitions between libration and circulation. This implies that the repeated separatrix crossing associated with the $(u,v)$ degree of freedom acts as the primary driver of stochasticity \citep{Yellowbook}. Moreover, recall that the equations of motion (\ref{eom}) dictate that the interactions between the degrees of freedom are facilitated exclusively by action coupling. Consequently, for further reduction of the model, it is sensible to parameterize the time-evolution of $\mathcal{W}$.

In absence of coupling, Hamiltonian (\ref{HAa}) describes simple pendulum-like dynamics for the $(\mathcal{W},w)$ degree of freedom. Under the assumption of small-amplitude libration of $w$ (which allows one to approximate the dynamics of a pendulum with that of a harmonic oscillator), $\mathcal{W}$ will exhibit sinusoidal variations \citep{Goldstein,MorbyBook}. Accordingly, let us adopt the following functional form:
\begin{align}
\mathcal{W} = \frac{[\Lambda]}{2} \left(\eta + \mu \cos(\sigma t) \right).
\label{Wt}
\end{align}
Empirically, the newly introduced constants and frequency are set to $\eta = 0.99$, $\mu = 0.025$, and $\sigma = 2\pi/(14.72)$ rad/year, to provide a suitable match to the numerical calculations. To this end, panel A of Figure (\ref{f5}) shows a comparison between the evolution of $\mathcal{W}$, computed within the framework of an $N$-body simulation (blue), numerical integration of the perturbative model governed by Hamiltonian (\ref{HAa}) (black) and the prescription (\ref{Wt}) (green). For consistency, a sinusoidal wave with the same frequency (corresponding to the envisioned evolution of $w$, given the approximation \ref{Wt}) is also depicted on panel B of Figure (\ref{f3}). We note that even though the adopted parameterization does not respect the physical requirement for the quantity $2\mathcal{W}/\Lambda$ to not exceed unity (see equation \ref{var2kepref}), it suffices for the purposes of an illustrative model we aim to construct.

Adopting equation (\ref{Wt}), equations of motion (\ref{eom}) can again be integrated numerically to yield an approximate dynamical evolution for the $(u,v)$ degree of freedom. The resulting solutions for the eccentricity as well as the critical angle $(2 \lambda - \lambda_{\rm{b}} + \gamma)$ are over-plotted as green curves in panels A and C of Figure (\ref{f3}) respectively. Although the shown solutions lose coherence (as chaotic systems must) after a few tens of years of evolution, the simplified model clearly captures the stochastic behavior exhibited by the system in a satisfactory manner. Thus, it can be sensibly employed to further characterize the dynamics in a rudimentary fashion. 

To understand the origins of chaos observed in the numerical solutions, note that under the assumption (\ref{Wt}), the equations of motion (\ref{eom}) correspond to a non-autonomous one degree of freedom Hamiltonian: 
\begin{align}
\tilde{\mathcal{H}} &=6 [n] \left(\frac{u^2 + v^2}{2} \right) + \beta u \nonumber \\
&- 6 [h] \left(\frac{[\Lambda]}{2} \left(\eta + \mu \cos(\sigma t) \right) + \frac{u^2 + v^2}{2} \right)^2 .
\label{HoneDOF}
\end{align}
This Hamiltonian possesses a single critical curve that sweeps across a region of phase-space with every circulation of the angle $\sigma t$ \citep{Henrard1982,Cary1986}. Accordingly, repeated encounters with the critical curve (which comprises an infinite-period orbit) render the motion on the separatrix-swept region of phase-space irregular \citep{Chirikov1979,Wisdom1985}.

Panel B of Figure (\ref{f5}) depicts the critical curve of Hamiltonian (\ref{HoneDOF}) at various phases of $\sigma t$. Shown on the same panel, is a stroboscopic surface of section arising from the same Hamiltonian, where the green points represent a chaotic sea while quasi-periodic trajectories are shown with gray curves. Note that as expected, the size of the chaotic region approximately conforms to the maximal phase-space area occupied by the separatrix. Indeed, this Figure is quite similar to the \Poincare\ surface of section shown in panel D of Figure (\ref{f4}). Moreover, the differences in the locations of families of quasi-periodic trajectories embedded in the chaotic sea, (shown in panels B, C and D of Figure \ref{f4}), can now be understood as changes in the extent of separatrix sweeping that result from alteration of the modulation at different values of $\mathcal{H}$.

It is noteworthy that the picture delineated in panel B of Figure (\ref{f5}) is not exactly one of adiabatic chaos \citep{Wisdom1985, Henrard1990}. This is made evident by the fact that quasi-periodic resonant trajectories that are not swept by the separatrix do not exist in this panel. This is likely due to the fact that the modulation frequency, $\sigma$, is so high, that the appearance and overlap of secondary resonances acts to wipe out this family of orbits \citep{Yellowbook}. For reference, panel C of Figure (\ref{f5}) shows a similar stroboscopic surface of section where the modulation frequency has been artificially reduced by a factor of 10. 

\subsection{Decoherence and Diffusion}

With a simplified picture at hand, we may now analytically derive the stochastic properties of the system. First and foremost, the above analysis allows us to obtain an estimate of the Lyapunov time. Crudely speaking, the Lyapunov time can be understood as a characteristic decoherence time of the system \citep{Yellowbook,MurrayHolman1997}. Accordingly, within the framework of Hamiltonian (\ref{HoneDOF}), it can be approximated as the time interval between successive encounters with the separatix, or half the modulation period:
\begin{align}
\tau_{\rm{L}} \sim \frac{1}{2} \left( \frac{2\pi}{\sigma} \right). 
\label{Lyap}
\end{align}
This simple functional form is in agreement with the estimates obtained by \citet{HolmanMurray1996} for the Asteroid belt. 

Given that the modulation period is slightly shorter than 15 years, the above relation suggests that the Lyapunov time associated with GJ876 ``e" should be of order $\tau_{\rm{L}}\sim 7$ years. To check this estimate numerically, we evaluated the Lyapunov time by integrating the full linearized variational equations (see \citealt{HolmanMurray1996}, Ch. 5 of \citealt{MorbyBook}, \citealt{Deck2013}) in parallel with a direct $N$-body simulation\footnote{The calculation was carried over $10^4$ years and the initial tangent vector was randomly oriented.}. The numerical calculation yielded a Lyapunov time of $\tau_{\rm{L}} = 7.26$ years, in excellent agreement with equation (\ref{Lyap}). As already mentioned above, this strongly suggests that in some similarity with the $Kepler$-36 system \citep{Deck2012}, the GJ876 system exhibits rapid dynamical chaos. 

Over timescales longer than a Lyapunov time, it is not sensible to treat any one trajectory as being representative. Instead, a statistical description of irregular trajectories is more appropriate \citep{Yellowbook,MurrayHolman1997}. In a uniformly chaotic region, the transport of actions obeys the Fokker-Plank equation \citep{Wang1945}, which reduces to the diffusion equation for Hamiltonian systems \citep{Landau1937}. Accordingly, the chaotic diffusion coefficient, $\mathcal{D}$, quantifies the essential attributes of dynamical evolution.

\begin{figure*}[t]
\centering
\includegraphics[width=1\textwidth]{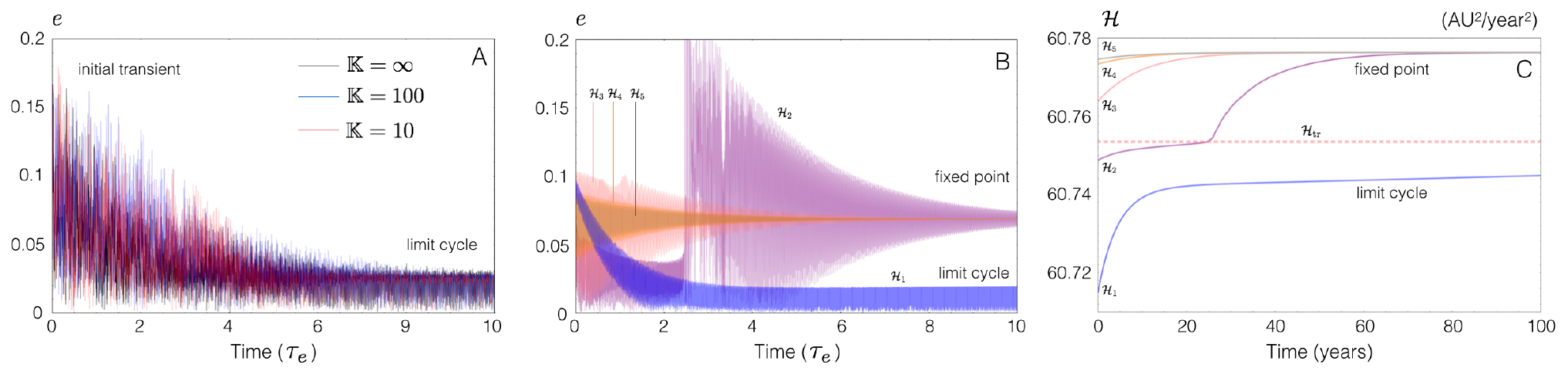}
\caption{Dissipative evolution of the multi-resonant system. Panel A shows the eccentricity of planet ``e" as a function of time obtained using dissipative $N$-body simulations. The behavior of the various curves is indistinguishable from each-other, as the results are not sensitive to the value of $\mathbb{K}$. Panel B shows the eccentricity evolution obtained within the framework of a dissipated perturbative model. Clearly, a quasi-steady limit cycle, such as that observed within the context of the $N$-body solution is one possible outcome observed in panel B. However, other evolutionary sequences that lead to an approach towards a fixed point are also viable. Panel C shows the dissipatively-facilitated evolution of $\mathcal{H}$. In this panel, the value of $\mathcal{H}$ corresponding to a transition from the limit-cycle attractor regime to the fixed-point attractor regime is shown with a dashed line.}
\label{f6}
\end{figure*}

In the quasi-linear approximation \citep{Murray1985}, the diffusion coefficient is given by the square of the typical change in action, $\Delta Z$, that takes place over a decoherence (i.e. Lyapunov) time, divided by the Lyapunov time. In the non-adiabatic regime (which is evidently representative of GJ876), correlations can be neglected (see \citealt{BruhwilerCary1989} for a discussion) and the trajectory can be envisioned to explore the chaotic layer uniformly. Accordingly, the average jump in action is of order half the size of the chaotic region. We have already mentioned that within the context of Hamiltonian (\ref{HoneDOF}), the extent of the chaotic layer approximately corresponds to the maximal phase-space area attained by the separatrix over a modulation cycle. Thus, a rough estimate for $\Delta Z$ is given by:
\begin{align}
\Delta Z \sim \frac{1}{2} \left(\frac{u_{\rm{hyp}}^2}{2}\right),
\label{DZ}
\end{align}
where $u_{\rm{hyp}}$ corresponds to the hyperbolic fixed point of the separatrix when $\sigma t = \pi$ (see Figure \ref{f5}) and is the solution to the equilibrium equation:
\begin{align}
0 =  \beta + 6 [n] u - 12 [h] u \left( \frac{[\Lambda]}{2} \left(\eta - \mu \right) + \frac{u^2}{2} \right).
\label{hyp}
\end{align}

Combining equations (\ref{Lyap}), (\ref{DZ}) and (\ref{hyp}), the explicit form of the diffusion coefficient becomes:
\begin{align}
\mathcal{D} &= \frac{\left(\Delta Z\right)^2}{\tau_{\rm{L}}} \sim \frac{\sigma}{16 \pi} \nonumber \\
&\times \big(2 (6)^{2/3} [h] ([n]+[h][\Lambda](\mu-\eta)) + (6)^{1/3}(3 [h]^2 \beta \nonumber \\ 
&+ \sqrt{3[h]^3(3[h] \beta^2-16([n]+[h][\Lambda](\mu-\eta))^3)} )^{2/3} \big)^4 \nonumber \\ 
&\times \big(\big( \sqrt{3[h]^3(3[h] \beta^2-16([n]+[h][\Lambda](\mu-\eta))^3)}\nonumber \\ 
&+3[h]^2 \beta \big)^{4/3}(1296 [h]^4) \big)^{-1}.
\label{DIFF}
\end{align}
Quantitatively, equation (\ref{DIFF}) evaluates to $\mathcal{D} \sim 3 \times 10^{-6} ([n] [\Lambda]^2)/(2\pi)$. This suggests that the diffusive progress of the eccentricity of planet ``e" over an orbital period is $\Delta e \sim \sqrt{3 \times 10^{-6}} \sim 10^{-3}$.

We can compare this result with a numerical estimate of the diffusion coefficient, calculated on a \Poincare\ surface of section. Specifically, we utilized the perturbative model (\ref{HAa}) to compute the average of the square of the change in action divided by the time-span between successive section intersections, using parameters corresponding to panel D of Figure (\ref{f4}). This procedure yielded $\mathcal{D}_{\rm{num}} = 6.15 \times 10^{-6} ([n] [\Lambda]^2)/(2\pi)$. Thus, the analytical estimate underestimates the numerically computed diffusion coefficient by a factor of $\sim 2$; an acceptable (and perhaps expected) error given the crudeness of the approximation involved in deriving equation (\ref{DIFF}).  

\section{Assembly of a Chaotic Laplace Resonance}

As already discussed in the introduction, assembly of the ``c-b" resonance has been studied extensively in the literature (\citealt{LeePeale2002,Crida2008} and the references therein). Had the discovery of the additional planet ``e" implied quasi-periodic behavior of the multi-resonant system (in some similarity to the case of the Galilean satellites), the assembly of the Laplace resonance would have been a straightforward extension of previous results \citep{Morby2007}. However, the chaotic nature of the orbits raises questions regarding the specific nature of the natal disk that is both sufficiently laminar to not preclude smooth migration (see e.g. \citealt{Bert2011}) and is simultaneously turbulent enough to prevent the system from settling to quasi-periodic depths of the resonant potential well \citep{Adams2008,Rein2009}.

For coherence, let us perform our investigation sequentially and first consider only dissipative effects. Accordingly, envisage the ``c-b-e" resonant chain embedded in a perfectly laminar protoplanetary nebula. Once locked in a mean-motion resonance, gap-opening giant planets (``b" and ``c") tend to carve out a vast mutual gap, greatly reducing the disk-facilitated migration rate \citep{MassetSnellgrove2001,MorbyCrida2007}. The same argument however does not apply to planet ``e", which likely experiences much stronger coupling with the disk. Thus, in line with the approximations invoked in the previous section, for the purposes of the perturbative model we can take the dynamics of the massive resonant pair to be isolated and periodic, while studying the dissipative evolution of the outermost planet in the test-particle limit.

\subsection{Dissipative Evolution}

Following \citet{LeePeale2002}, we shall parameterize the effects of planet-disk interaction using the following simple relationships:
\begin{align}
\frac{de}{dt} = - \frac{e}{\tau_e} \ \ \ \ \ \ \ \ \frac{da}{dt} = - \frac{1}{\mathbb{K}} \frac{a}{\tau_e},
\label{Nebula}
\end{align}
where $\tau_e$ is the characteristic eccentricity damping timescale, and $\mathbb{K}$ is a constant that parameterizes the semi-major axis damping timescale in terms of $\tau_e$. Because we are interested in understanding the long-term evolution of a chaotic multi-resonant state within the disk, we shall adopt the present (observed) state as an adequate initial condition. In line with the approximations quoted above, we have performed a series of $N$-body simulations where only the outermost planet in the system is affected by the fictitious dissipative forces\footnote{Note that this differs from the analysis of e.g. \citet{LeePeale2002}, who initialized the system in a non-resonant state to study the capture process.} (\ref{Nebula}). 

Recall that the resonant modulation time invoked in the previous section is $2\pi/\sigma \sim 15$ years, much shorter than the typically quoted estimates for $\tau_e$. Therefore, in accordance with adiabatic theory \citep{Henrard1982}, the final outcome of our simulations is insensitive to the exact value of $\tau_e$ (which we safely set to $\tau_e = 10^3$ years). We note further that the adopted adiabatic regime\footnote{In this limit, $\tau_e$ can be used to replace the effective unit of time associated with dissipative evolution, allowing the obtained results to be scaled to other values of $\tau_e$.} is consistent with the observed orbital state, since resonant capture probabilities diminish significantly in non-adiabatic systems.

Panel A of Figure (\ref{f6}) shows the eccentricity evolution of planet ``e" observed in the dissipated $N$-body simulations. Specifically, the calculations suggest that following an initial transient period of order $\sim 5 \tau_e$, the system settles onto a quasi-periodic limit cycle. Moreover, the evolutionary sequence appears to be independent of the assumed value of $\mathbb{K}$, as long as it exceeds $\mathbb{K} \gtrsim 5$. To this end, we note that \citet{LeePeale2002} favor a value of $\mathbb{K} \sim 100$ for the natal disk of GJ876 (which we adopt for subsequent calculations), rendering our results rather robust.

As with chaotic diffusion, the quoted results of dissipative $N$-body simulations can be understood within the framework of the perturbative model delineated in the previous section. That said, care must be taken in implementing equations (\ref{Nebula}). Explicitly, the definition of variables (\ref{finangles}) implies that both degrees of freedom will be affected by dissipative evolution. However, the perturbative model differs from the $N$-body system in a crucial manner: the test particle approximation employed in the former impedes resonant transport of the chain. That is, although realistically the application of semi-major axis migration to planet ``e" affects the radial evolution of the whole system, this effect is not captured in the formulation of the Hamiltonian (\ref{HAa})\footnote{An explicit damping of $\mathcal{W}$ introduces an unphysical dependence on $\mathbb{K}$ into the perturbative model.}. Consequently, given that the action $\mathcal{W}$ is directly proportional to $\Lambda$, a more physical representation of the real system can be attained by only applying dissipative effects to the $(u,v)$ degree of freedom. Suitably, the revised equations of motion take the form:
\begin{align}
\frac{du}{dt} &= \left( \frac{du}{dt} \right)_{\mathcal{H}} - \frac{u}{\tau_e} \nonumber \\
\frac{dv}{dt} &= \left( \frac{dv}{dt} \right)_{\mathcal{H}} - \frac{v}{\tau_e},
\label{eomuvdiss}
\end{align}
where the subscript $\mathcal{H}$ signifies the Hamiltonian contribution. The equations corresponding to the ($\mathcal{W},w$) degree of freedom remain unchanged from (\ref{eom}).

Using the modified perturbative model, we have simulated the dissipative evolution of the system, starting with different values of $\mathcal{H}$ depicted in Figure (\ref{f4}). The calculated eccentricity is shown as a function of time in panel B of Figure (\ref{f6}), where each curve is labeled by the starting value of the Hamiltonian. The corresponding evolution of $\mathcal{H}$ is shown in panel C of Figure (\ref{f6}). Evidently, two distinct modes of dissipative evolution exist: the approach to a limit cycle (as shown by the blue curve, $\mathcal{H}_1$) and the approach towards a fixed point.

As long as the damping is comparatively slow (as discussed above), at each point along the evolutionary track, a purely Hamiltonian snap-shot of the dynamical portrait yields a good representation of the orbital state. Thus, the dichotomy inherent to the behavior of the dissipative evolution can be understood by examining the surfaces of section depicted in Figure (\ref{f4}). The addition of dissipative forces into the model alters the dynamical behavior into two ways: it leads to a gradual reduction of the phase-space area enclosed by a given orbit in phase-space (see e.g. \citealt{Yoder1973,Yoder1979,BatyginMorbielli2011} for a related discussion) and an increase in the value of the Hamiltonian. Consequently, if the orbit is initialized somewhere within the chaotic layer (see panels B, C and D of Figure \ref{f4}), as time proceeds the orbit will tend to exit the chaotic sea and settle onto the center of the corresponding quasi-periodic island. Simultaneously, the dynamical portraits will change in such a manner that the area enclosed by the constant-energy curves on the double section (panel A of Figure \ref{f4}) will also decrease. The two processes are interrelated as the rate of change of $\mathcal{H}$ grows as the action $\mathcal{Z}$ (associated with the $(u,v)$ degree of freedom) increases. 

If the starting condition of a $\mathcal{H} = \mathcal{H}_1$ orbit is chosen such that it is relatively close to the center of the quasi-periodic island depicted in panel B of Figure (\ref{f4}), the orbit will rapidly decay onto a limit-cycle that intersects the center of the associated \Poincare\ surface of section. Concurrently, because such a limit cycle is characterized by a consistently low eccentricity, the rate of dissipative increase of $\mathcal{H}$ will be reduced, rendering such a state of the system quasi-steady. An example of such an evolution is shown by the blue curves in panels B and C of Figure (\ref{f6}). The $N$-body solution shown in panel A of Figure (\ref{f6}) also exhibits such behavior.

If the system is initialized at a higher level of $\mathcal{H}$ (e.g. $\mathcal{H}_3$, $\mathcal{H}_4$, $\mathcal{H}_5$), the corresponding \Poincare\ surfaces of section depicted in Figure (\ref{f4}) show that the limit cycle (which turns out to correspond to a fixed point) resides at a high eccentricity meaning that the evolution of $\mathcal{H}$ will also proceed at an unhindered rate. Consequently, on a timescale not much grater that $\sim \tau_e$, the system will settle onto an equilibrium described by a maximal attainable value of $\mathcal{H}$ (which also corresponds to a null area enclosed by the orbit in the double-section shown on Figure \ref{f4}). Solutions of this kind are shown as pink ($\mathcal{H}_3$), orange ($\mathcal{H}_4$), and brown ($\mathcal{H}_5$) curves on panels B and C of Figure (\ref{f6}).

The value of $\mathcal{H}$ that separates the two regimes lies between $\mathcal{H}_2 < \mathcal{H}_{\rm{tr}} < \mathcal{H}_3$. Thus, it is possible to have a trajectory that first evolves onto a limit cycle, but upon crossing the critical value of $\mathcal{H}$, breaks out of the limit cycle\footnote{This happens because the associated island of stability gets engulfed by the chaotic sea.} and begins its approach to the fixed point. An example of such an evolution is shown in purple ($\mathcal{H}_2$) on panels B and C of Figure (\ref{f6}).

Irrespective of the details of the solution, a common feature observed in all evolutionary tracks is the approach to quasi-periodicity on a timescale comparable to $\tau_e$. Numerical simulations (see e.g. \citealt{Ogilvie2003,Papaloizou2007}) suggest that $\tau_e$ is considerably shorter than the lifetime of a disk. Consequently, in absence of additional perturbations, the formation of a chaotic Laplace resonance in a perfectly laminar protoplanetary disk appears unlikely. Accordingly, in the following discussion we shall invoke turbulent forcing as a means of preventing the system from reaching dynamical equilibration.

\subsection{Damped, Driven Evolution}

\begin{figure*}[t]
\centering
\includegraphics[width=1\textwidth]{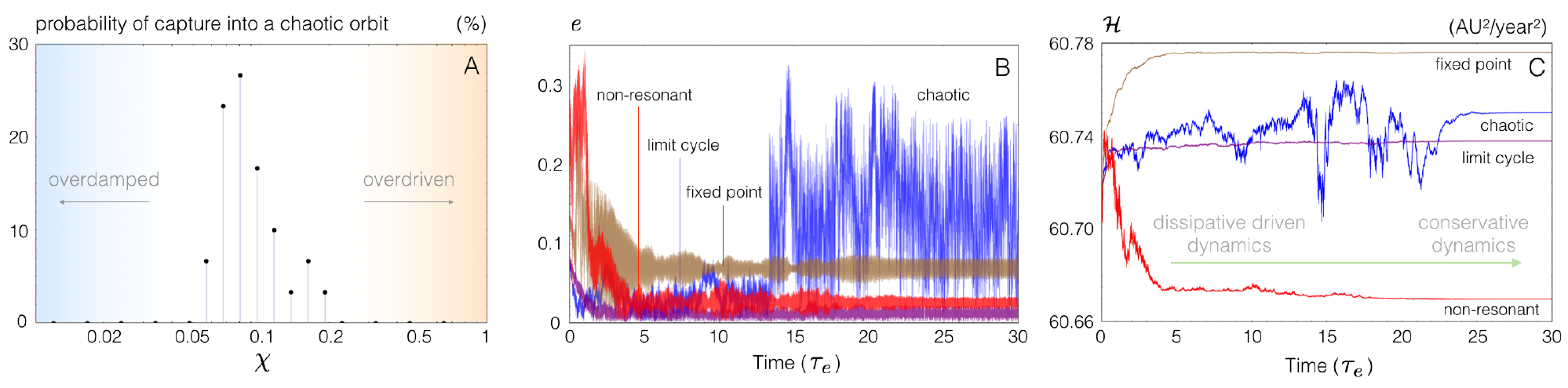}
\caption{Stochastically driven, dissipative evolution of the multi-resonant chain. Panel A shows the probability of a chaotic end-state as a function of the dimensionless parameter $\chi$. Panels B and C depict four representative solutions observed in our Monte-Carlo simulations, and parallel panels B and C of Figure (6). Specifically, orbits that approach a chaotic end state (blue), a limit cycle (purple), a fixed point (green) as well as a non-resonant state (red) are shown. Recall that the calculations are performed in such a way that the stochastic forcing as well as dissipative effects diminish with time, leading to a nearly conservative system at the end of the integrations. }
\label{f7}
\end{figure*}

Angular momentum transport within protoplanetary disks (that facilitates accretion of disk material onto the host stars) is typically attributed to turbulence \citep{Armitage2011} that is expected to stem from the magneto-rotational instability \citep{BalbusHawley1991} or some other process. Turbulent effects can have important consequences for resonant coupling and have generally been shown to lead to a breakup of resonant libration \citep{Adams2008,Rein2009,Ketchum2011}. Therefore, provided sufficiently vigorous stochastic diffusion, it is reasonable to expect that turbulent perturbations may overcome dissipative interactions, allowing for the formation of a chaotic multi-resonant chain.

As is with laminar disk-star interactions, the most direct way to calculate the evolution of embedded planets is using global 3D magnetohydrodynamic simulations (see for example \citealt{FromangNelson2006}). However, such simulations can be computationally expensive and will likely preclude a statistically sound exploration of the relevant parameter regime. As a result, here we shall again opt for a parametrized treatment of stochastic forcing and perform the simulations within the framework of the perturbative model. 

The simplest approach to mimicking stochastic forcing is by introducing Gaussian white noise, $\mathscr{F}$, into the equations of motion \citep{Adams2008,Rein2009}. Following the purely dissipative treatment outlined above, we shall only apply these effects to the $(u,v)$ degree of freedom. Moreover, as we will seek to examine the evolved outcome of these simulations, we shall additionally introduce a time-dependence, $\Upsilon$, to the damping and driving terms which will (over a time considerably greater than $\tau_e$) cause these effects to vanish slowly. Accordingly, the relevant equations of motion become: 
\begin{align}
\frac{du}{dt} &= \left( \frac{du}{dt} \right)_{\mathcal{H}} + \left[ \mathscr{F}_u - \frac{u}{\tau_e} \right] \Upsilon \nonumber \\
\frac{dv}{dt} &= \left( \frac{dv}{dt} \right)_{\mathcal{H}} + \left[ \mathscr{F}_v - \frac{v}{\tau_e}  \right] \Upsilon.
\label{dampdriven}
\end{align}

Equations of the form (\ref{dampdriven}) are typically referred to as Langevin equations and yield solutions that should be interpreted in a statistical sense \citep{Stochastic}. It is well known that the integration of white noise, $\mathscr{F}$, yields the Weiner process, $\mathscr{W}$ (a scaling limit of a random walk). With a zero expectation value, the progress of a pure Weiner process is $\sim \sqrt{\mathscr{D} t}$, where the $\mathscr{D}$ is the effective diffusion coefficient. $\mathscr{D}$ is directly related to the variance of $\mathscr{F}$, which we take as an adjustable parameter.

If the stochastic term in the equations of motion is augmented with dissipative effects (i.e. the terms within the brackets of equations \ref{dampdriven}), the progress of the associated quantity becomes limited from above and upon saturation, approaches $\sim \sqrt{\mathscr{D} \tau_e}$. Consequently, for the system at hand, we can define a dimensionless number:
\begin{align}
\chi \equiv \sqrt{\frac{\mathscr{D} \tau_e}{[\Lambda]} } \leqslant 1,
\label{dimentsionless}
\end{align}
that approximately characterizes the maximal eccentricity that the orbit would attain exclusively due to interactions with the nebula\footnote{Because of this interpretation, it is sensible to limit $\chi$ by unity from above, for bound systems.}. Note that the physical meaning of $\chi$ roughly parallels that of  the turbulent Schmidt number: it represents a ratio of turbulent forcing to viscous dissipation. Thus, a value of $\chi$ that leads to the formation of a system that resembles the real GJ876 resonant chain is informative of the properties of the protoplanetay disk in which the system was assembled.

Of course, the dynamical behavior of the actual damped, driven two degree of freedom system is quite complex (in part because the value of $\mathcal{H}$ changes substantially), and it is not clear what value of $\chi$ leads to favorable evolution a-priori. Thus, we have performed a series of Monte-Carlo numerical experiments in an effort to survey the characteristic outcome as a function of $\chi$. For all simulations, a starting value of $\mathcal{H} = \mathcal{H}_1$, corresponding to the \Poincare\ surface of section depicted in panel B of Figure (\ref{f4}), was adopted and the initial condition was chosen randomly on the section. Indeed, the choice of the starting value of $\mathcal{H}$ is somewhat arbitrary and is generally unimportant because (with the exception of a finely tuned set of parameters) the system quickly loses memory of its starting state.

Each integration spanned $30\, \tau_e = 3 \times 10^4$ years and the functional form for the time-dependence quoted in equations (\ref{dampdriven}) was chosen to be $\Upsilon = (\exp(-t/\bar{\tau}))^{6}$, where $\bar{\tau} = 20\, \tau_e$. A range of $10^{-2} \leqslant \chi \leqslant 1$ was explored and 30 realizations of the system were integrated per choice of $\chi$. Upon completion of the integrations, the end state of each simulation was used as an initial condition for a purely Hamiltonian integration (equations \ref{eom}) and a surface of section corresponding to the particular orbit was constructed. The stochastic properties of the orbit were then examined. 

The results of the Monte-Carlo survey are shown in Figure (\ref{f7}). Most importantly, panel A shows the probability of capture into a chaotic Laplace resonance as a function of $\chi$. It is striking that the probability of reproducing the observed state can be substantial within the context of our model, however this remains the case only for a rather restricted range of parameters. Particularly, the probability of success is nearly $30\%$ for $\chi \simeq 8 \times 10^{-2}$ but drops to zero for $\chi \lesssim 5 \times 10^{-2}$ and $\chi \gtrsim 2 \times 10^{-1}$. 

The reason for a relatively narrow range of $\chi$ that allows for chaotic resonances as an outcome has to do with the evolution of $\mathcal{H}$. In the case of the over-damped system (where $\chi$ is low), the evolution essentially proceeds as described in the preceding discussion of a purely dissipative setup. That is, the trajectory collapses onto a nearby quasi-periodic island and $\mathcal{H}$ evolves to a stationary value characterized by an equilibrated orbital state. In the case of an over-driven system, a somewhat different picture emerges. As turbulent forcing allows the orbit to explore the phase-space stochastically, the system eventually exits the resonance all together, and settles onto an orbit dominated by secular interactions (not accounted for in our model). Such evolutions are marked by a substantial decrease in the value of the Hamiltonian. 

The evolution of the eccentricity and the value of $\mathcal{H}$ for representative trajectories observed in our simulations are plotted in panels B and C of Figure (\ref{f7}) respectively. The evolutionary track shown in brown, exhibits some similarity to that discussed within the context of purely dissipative evolution. Specifically, $\mathcal{H}$ increases to its maximal value and the system settles to the vicinity of a fixed point. Note however, that owing to turbulent forcing, the system retains a finite libration amplitude at the end of the simulation, shown by non-negligible oscillation of the eccentricity. The purple evolutionary track is also similar to the purely dissipative case, but lies in the limit-cycle category. Correspondingly, the eccentricity evolution resembles the results of $N$-body integrations, depicted in panel A of Figure (\ref{f6}). 

An over-driven simulation, where the system breaks out of resonance and ultimately settles onto a quasi-periodic orbit (see also \citealt{Adams2008}) is shown in red. Note that in this simulation, the evolution of the Hamiltonian occurs in the opposite sense compared to the over-damped simulation. That is, the value of $\mathcal{H}$ decreases relative to its initial condition. Finally, an orbit that settles onto a chaotic Laplace resonance is shown in blue. Here, the value of $\mathcal{H}$ remains close to the $\mathcal{H}_1 - \mathcal{H}_3$ range where the \Poincare\ surfaces of section depicted in Figure (\ref{f4}) show a considerable fraction of phase-space that is occupied by chaotic seas. As such, in this simulation the system successfully acquires a chaotic end-state, facilitated by stochastic elimination of stable quasi-periodic trajectories throughout the evolutionary sequence.

\section{Discussion}

In this paper, we have examined the dynamical state as well as the formation scenario of a multi-resonant chain of planets residing in the GJ876 system. In particular, we began our investigation by constructing a simplified model, based on Hamiltonian perturbation theory, that broadly captures the dynamical behavior of the system. Upon a detailed examination of this model, we considered the assembly of the multi-resonant system in presence of dissipative and stochastic forces. 

With the tally of exoplanetary detections now in excess of a thousand \citep{Wright2011,Batalha2013}, it is tempting to question the value held in detailed examination of the orbital architecture of a particular system. While such criticism may be appropriate for bodies whose overall state is reminiscent of other well-characterized exoplanets, the GJ876 system easily escapes such scrutiny as it clearly stands out as a unique member of the observational aggregate. To this end, the GJ876 ``c-b-e" system represents the only known extrasolar Laplace-like resonance. Unlike the the Galilean satellites however, the GJ876 multi-resonant system is vigorously chaotic, with a characteristic decoherence time on the order of a decade. The primary difference between the two configurations stems from the high eccentricities attained by the massive planets in the GJ876 system and the fact that the resonant libration amplitudes associated with planet ``e" never approached near-null values. While interesting in its own right, the nature of the GJ876 resonance plays an important additional role as a means of placing much-needed constraints on the nature of the protoplanetary disk from which the system emerged. 

The characterization of the stability, dynamical structure, and origins of the Laplace resonance inherent to the Galilean satellites (within a tangible framework) is among the most widely-celebrated achievements of celestial mechanics of the latter half of the 20th century (see \citealt{Peale1976,Peale1986} for reviews). As we have shown here, the fiercely chaotic Laplace-like resonance of the GJ876 system can also be understood within the context of a simple time-independent perturbative model, characterized by two degrees of freedom. Specifically, the mathematical formulation of the governing Hamiltonian is analogous to a pair of momentum-coupled pendulums. While in agreement with direct numerical integration (\citealt{Rivera2010}), this model clearly illuminates the origins of stochastic motion and allows one to derive simple analytic estimates of the Lyapunov time and action diffusion coefficient related to the outer-most planet's eccentricity \citep{Yellowbook,MurrayHolman1997}. 

Given the system's short ($\sim$ decadal) timescale for stochastic excursions (implied by the analytical estimates and observed in the numerical simulations), one may readily expect the associated quasi-random signal to become evident in the observational radial velocity time series of the system. Specifically, one may naively expect that the individual lines observed on the periodogram will be chaotically broadened \citep{Laskar1993b}. This consideration may place fundamental limits on any dynamical model's ability to match the observed signal, even provided an outstanding signal to noise ratio (see \citealt{LaughlinChambers2001,Rivera2010,Deck2012}). In practice, however, the chaotic nature of the orbits is probably not the dominant source of error in the data \citep{Baluev2011} and likely leads to a limited reduction in the goodness of fit \citep{Gregchat}. 

A crucial feature that arises naturally within the context of the developed framework is the simultaneous rapidity of the chaotic loss of dynamical memory and long-term stability. This characteristic contrasts the de-populated mean motion resonances (Kirkwood gaps) within the Asteroid belt \citep{Wisdom1983,Wisdom1985,Murray1986,Henrard1990,Morby1990a,Morby1990b,NesvornyMorbidelli1998} but bears some resemblance to the long-term stability of the planets of the Solar System \citep{Laskar1989,Laskar1996,SussmanWisdom1988,SussmanWisdom1992,Quinn1991,MurrayHolman1999}. The analogy is in fact surprisingly robust, especially for the case of Mercury. That is, much like GJ876 ``e," Mercury is characterized by a Lyapunov time that is orders of magnitude shorter than the Solar System's lifetime and has only a slim chance of escaping the Solar System within the remaining main-sequence lifetime of the Sun \citep{Laskar2008,BatyginLaughlin2008,Laskar2009}. In a related effort, \citet{LithWu11} have shown that Mercury's secular evolution can be understood within the context of an autonomous Hamiltonian corresponding to a pair of momentum-coupled pendulums\footnote{Aside from distinct meanings of the variables, a slight difference between the perturbative representations of Mercury and GJ876 ``e" is that both pendulums posses the D'Almbert characteristic \citep{HenrardLemaitre1983} in the former case, while one of the pendulums is simple in the latter case.} (see also \citealt{Sid1990,Boue2012}), much like the case of the resonant evolution of GJ876 ``e". Furthermore, in direct correspondence to the stability of GJ876, \citet{Batyginetal2014} have recently shown that Mercury's long-term stability also arises from a topological boundary associated with the approximate conservation of the Hamiltonian itself (see \Poincare\ surfaces of section depicted in Figure \ref{f4}).

The formation of the GJ876 multi-resonant chain almost certainly requires convergent approach of the orbits facilitated disk-driven migration. While a perfectly laminar evolutionary track had previously been invoked to explain the (nearly periodic) ``c-b" mean motion resonance (see \citealt{LeePeale2002} as well numerous other references quoted above), our modeling suggests that a chaotic configuration is incompatible with formation in a purely dissipative nebular environment. Instead, limited stochastic perturbations arising from turbulent forcing \citep{Adams2008,Rein2009} appear to be needed to explain the observed configuration. Accordingly, we utilized our perturbative model to survey a range of assumed disk parameters with an eye towards identifying a regime that leads to a successful formation of a chaotic multi-resonant chain. This allowed us to statistically infer (albeit within the framework of an approximate model) the likely combination of relative strengths of turbulent perturbations and dissipative effects inherent to the GJ876 protoplanetary disk.

The corresponding dimensionless quantity derived in section (4.2) is related to the turbulent Schmidt number of the disk. Although parameters of this sort are of great importance for the quantification of angular momentum transport within disks \citep{Armitage2011}, their observational characterization is at present scarce \citep{Hughes2011}. Thus, theoretical considerations such as that undertaken in this work are required to inform the relevant characteristics. Systems such as GJ876 provide a rare opportunity to perform such an analysis. Accordingly, a viable and consequential extension of the presented work would involve direct magnetohydrodynamic modeling of the assembly of the GJ876 multi resonant system within its natal disk (see e.g. \citealt{Nelson2005,NelsonGressel2010}). Such an investigation would no doubt further our understanding of typical formation environments of planetary systems and place additional constraints on the dominant processes involved in sculpting their orbital architectures.  

\acknowledgments
\textbf{Acknowledgments}. We are grateful to Greg Laughlin and Fred Adams for useful discussions.


\begin{thebibliography} 

\bibitem[Adams et al.(2008)]{Adams2008} Adams, F.~C., Laughlin, G., \& Bloch, A.~M.\ 2008, \apj, 683, 1117 

\bibitem[Armitage(2011)]{Armitage2011} Armitage, P.~J.\ 2011, \araa, 49, 195 

\bibitem[Balbus \& Hawley(1991)]{BalbusHawley1991} Balbus, S.~A., \& Hawley, J.~F.\ 1991, \apj, 376, 214 

\bibitem[Baluev(2011)]{Baluev2011} Baluev, R.~V.\ 2011, Celestial Mechanics and Dynamical Astronomy, 111, 235 

\bibitem[Batalha et al.(2013)]{Batalha2013} Batalha, N.~M., Rowe, J.~F., Bryson, S.~T., et al.\ 2013, \apjs, 204, 24 

\bibitem[Batygin \& Laughlin(2008)]{BatyginLaughlin2008} Batygin, K., \& Laughlin, G.\ 2008, \apj, 683, 1207 

\bibitem[Batygin \& Brown(2010)]{BatyginBrown2010} Batygin, K., \& Brown, M.~E.\ 2010, \apj, 716, 1323 

\bibitem[Batygin et al.(2011)]{BatyginBrownFraser2011} Batygin, K., Brown, M.~E., \& Fraser, W.~C.\ 2011, \apj, 738, 13 

\bibitem[Batygin \& Morbidelli(2011)]{BatyginMorbielli2011} Batygin, K., \& Morbidelli, A.\ 2011, Celestial Mechanics and Dynamical Astronomy, 111, 219

\bibitem[Batygin et al.(2012)]{BatyginBrownBetts2012} Batygin, K., Brown, M.~E., \& Betts, H.\ 2012, \apjl, 744, L3 

\bibitem[Batygin(2012)]{Batygin2012} Batygin, K.\ 2012, \nat, 491, 418 

\bibitem[Batygin \& Morbidelli(2013)]{BatyginMorbidelli2013} Batygin, K., \& Morbidelli, A.\ 2013, \aap, 556, A28

\bibitem[Batygin et al.(2014)]{Batyginetal2014} Batygin, K., Morbidelli, A., \& Holman, M., 2014, \apj, \rm{submitted}

\bibitem[Beaug{\'e} \& Nesvorn{\'y}(2012)]{BeaugeNesvorny2012} Beaug{\'e}, C., \& Nesvorn{\'y}, D.\ 2012, \apj, 751, 119 

\bibitem[Bean \& Seifahrt(2009)]{BeanSeifahrt2009} Bean, J.~L., \& Seifahrt, A.\ 2009, \aap, 496, 249 

\bibitem[Beaug{\'e} \& Michtchenko(2003)]{Beauge2003} Beaug{\'e}, C., \& Michtchenko, T.~A.\ 2003, \mnras, 341, 760 

\bibitem[Beaug{\'e} et al.(2003)]{Beaugeetal2003} Beaug{\'e}, C., Ferraz-Mello, S., \& Michtchenko, T.~A.\ 2003, \apj, 593, 1124 

\bibitem[Beaug{\'e} et al.(2006)]{Beaugeetal2006} Beaug{\'e}, C., Michtchenko, T.~A., \& Ferraz-Mello, S.\ 2006, \mnras, 365, 1160 

\bibitem[Benedict et al.(2002)]{Benedict2002} Benedict, G.~F., McArthur, B.~E., Forveille, T., et al.\ 2002, \apjl, 581, L115 

\bibitem[Bitsch \& Kley(2011)]{Bert2011} Bitsch, B., \& Kley, W.\ 2011, \aap, 536, A77 

\bibitem[Bitsch et al.(2013)]{Bert2013} Bitsch, B., Boley, A., \& Kley, W.\ 2013, \aap, 550, A52 

\bibitem[Bou{\'e} et al.(2012)]{Boue2012} Bou{\'e}, G., Laskar, J., \& Farago, F.\ 2012, \aap, 548, A43 

\bibitem[Bruhwiler \& Cary(1989)]{BruhwilerCary1989} Bruhwiler, D.~L., \& Cary, J.~R.\ 1989, Physica D Nonlinear Phenomena, 40, 265 

\bibitem[Cary et al.(1986)]{Cary1986} Cary, J.~R., Escande, D.~F., \& Tennyson, J.~L.\ 1986, \pra, 34, 4256 

\bibitem[Chambers(1999)]{Chambers1999} Chambers, J.~E.\ 1999, \mnras, 304, 793 

\bibitem[Chirikov(1979)]{Chirikov1979} Chirikov, B.~V.\ 1979, \physrep, 52, 263 

\bibitem[Correia et al.(2010)]{Correia2010} Correia, A.~C.~M., Couetdic, J., Laskar, J., et al.\ 2010, \aap, 511, A21 

\bibitem[Crida et al.(2007)]{Crida2007} Crida, A., Morbidelli, A., \& Masset, F.\ 2007, \aap, 461, 1173 

\bibitem[Crida et al.(2008)]{Crida2008} Crida, A., S{\'a}ndor, Z., \& Kley, W.\ 2008, \aap, 483, 325 

\bibitem[Deck et al.(2012)]{Deck2012} Deck, K.~M., Holman, M.~J., Agol, E., et al.\ 2012, \apjl, 755, L21 

\bibitem[Deck et al.(2013)]{Deck2013} Deck, K.~M., Payne, M., \& Holman, M.~J.\ 2013, \apj, 774, 129

\bibitem[Delfosse et al.(1998)]{Delfosse1998} Delfosse, X., Forveille, T., Mayor, M., et al.\ 1998, \aap, 338, L67 

\bibitem[Delisle et al.(2014)]{Delisle2014} Delisle, J.-B., Laskar, J., \& Correia, A.~C.~M.\ 2014, arXiv:1404.4861

\bibitem[Ellis \& Murray(2000)]{EllisMurray2000} Ellis, K.~M., \& Murray, C.~D.\ 2000, \icarus, 147, 129

\bibitem[Escande(1985)]{Escande1985} Escande, D.~F.\ 1985, \physrep, 121, 165 

\bibitem[Fabrycky et al.(2014)]{Fabrycky2012} Fabrycky, D.~C., Lissauer, J.~J., Ragozzine, D., et al.\ 2014, \apj, 790, 146

\bibitem[Fromang \& Nelson(2006)]{FromangNelson2006} Fromang, S., \& Nelson, R.~P.\ 2006, \aap, 457, 343

\bibitem[Gerlach \& Haghighipour(2012)]{Gerlach2012} Gerlach, E., \& Haghighipour, N.\ 2012, Celestial Mechanics and Dynamical Astronomy, 113, 35 

\bibitem[Goldreich(1965)]{Goldreich1965} Goldreich, P.\ 1965, \mnras, 130, 159

\bibitem[Goldreich \& Tremaine(1980)]{GoldreichTremaine1980} Goldreich, P., \& Tremaine, S.\ 1980, \apj, 241, 425 

\bibitem[Goldstein(1950)]{Goldstein} Goldstein, H.\ 1950, Classical Mechanics, (Mass.: Addison-Wesley)  

\bibitem[Go{\'z}dziewski et al.(2002)]{Goz2002} Go{\'z}dziewski, K., Bois, E., \& Maciejewski, A.~J.\ 2002, \mnras, 332, 839 

\bibitem[Greenberg(1977)]{Greenberg1977} Greenberg, R.\ 1977, \baas, 9, 520 

\bibitem[Haghighipour et al.(2003)]{Nader2003} Haghighipour, N., Couetdic, J., Varadi, F., \& Moore, W.~B.\ 2003, \apj, 596, 1332 

\bibitem[Henrard(1982)]{Henrard1982} Henrard, J.\ 1982, Celestial Mechanics, 27, 3 

\bibitem[Henrard \& Lemaitre(1983)]{HenrardLemaitre1983} Henrard, J., \& Lemaitre, A.\ 1983, \icarus, 55, 482 

\bibitem[Henrard et al.(1986)]{Henrard1986} Henrard, J., Milani, A., Murray, C.~D., \& Lemaitre, A.\ 1986, Celestial Mechanics, 38, 335 

\bibitem[Henrard \& Caranicolas(1990)]{Henrard1990} Henrard, J., \& Caranicolas, N.~D.\ 1990, Celestial Mechanics and Dynamical Astronomy, 47, 99

\bibitem[Holman \& Murray(1996)]{HolmanMurray1996} Holman, M.~J., \& Murray, N.~W.\ 1996, \aj, 112, 1278 

\bibitem[Hughes et al.(2011)]{Hughes2011} Hughes, A.~M., Wilner, D.~J., Andrews, S.~M., Qi, C., \& Hogerheijde, M.~R.\ 2011, \apj, 727, 85 

\bibitem[Ji et al.(2002)]{Ji2002} Ji, J., Li, G., \& Liu, L.\ 2002, \apj, 572, 1041 

\bibitem[Jones et al.(2001)]{Jones2001} Jones, B.~W., Sleep, P.~N., \& Chambers, J.~E.\ 2001, \aap, 366, 254 

\bibitem[Juri{\'c} \& Tremaine(2008)]{JuricTremaine2008} Juri{\'c}, M., \& Tremaine, S.\ 2008, \apj, 686, 603 

\bibitem[Ketchum et al.(2011)]{Ketchum2011} Ketchum, J.~A., Adams, F.~C., \& Bloch, A.~M.\ 2011, \apj, 726, 53 

\bibitem[Klebaner(2012)]{Stochastic} Klebaner, F. C., 1998, Introduction to Stochastic Calculus with Applications (World Scientific Publishing)

\bibitem[Kley et al.(2004)]{Kley2004} Kley, W., Peitz, J., \& Bryden, G.\ 2004, \aap, 414, 735 

\bibitem[Kley et al.(2005)]{Kley2005} Kley, W., Lee, M.~H., Murray, N., \& Peale, S.~J.\ 2005, \aap, 437, 727

\bibitem[Kley \& Nelson(2012)]{KleyNelson2012} Kley, W., \& Nelson, R.~P.\ 2012, \araa, 50, 211

\bibitem[Kinoshita \& Nakai(2001)]{KinoshitaNakai2001} Kinoshita, H., \& Nakai, H.\ 2001, \pasj, 53, L25 

\bibitem[Landau(1937)]{Landau1937} Landau, L.\ 1937, Physical Review, 52, 1251 

\bibitem[Laughlin \& Chambers(2001)]{LaughlinChambers2001} Laughlin, G., \& Chambers, J.~E.\ 2001, \apjl, 551, L109 

\bibitem[Laughlin(private communication)]{Gregchat} Laughlin, G., 2014, private communication

\bibitem[Laskar(1989)]{Laskar1989} Laskar, J.\ 1989, \nat, 338, 237 

\bibitem[Laskar(1993a)]{Laskar1993} Laskar, J.\ 1993a, Celestial Mechanics and Dynamical Astronomy, 56, 191 

\bibitem[Laskar(1993b)]{Laskar1993b} Laskar, J.\ 1993b, Physica D Nonlinear Phenomena, 67, 257 

\bibitem[Laskar(1996)]{Laskar1996} Laskar, J.\ 1996, Celestial Mechanics and Dynamical Astronomy, 64, 115 

\bibitem[Laskar(2008)]{Laskar2008} Laskar, J.\ 2008, \icarus, 196, 1 

\bibitem[Laskar \& Gastineau(2009)]{Laskar2009} Laskar, J., \& Gastineau, M.\ 2009, \nat, 459, 817 

\bibitem[Lee \& Peale(2002)]{LeePeale2002} Lee, M.~H., \& Peale, S.~J.\ 2002, \apj, 567, 596 

\bibitem[Lee(2004)]{Lee2004} Lee, M.~H.\ 2004, \apj, 611, 517 

\bibitem[Lee \& Thommes(2009)]{LeeThommes2009} Lee, M.~H., \& Thommes, E.~W.\ 2009, \apj, 702, 1662 

\bibitem[Lega et al.(2013)]{Lega2013} Lega, E., Morbidelli, A., \& Nesvorn{\'y}, D.\ 2013, \mnras, 431, 3494 

\bibitem[Leverrier(1855)]{LeVerrier1855} Leverrier, U. J.\ 1855, Annales de llObservatoire de Paris, 1, 258

\bibitem[Levison et al.(2008)]{Levison2008} Levison, H.~F., Morbidelli, A., Van Laerhoven, C., Gomes, R., \& Tsiganis, K.\ 2008, \icarus, 196, 258 

\bibitem[Lichtenberg \& Lieberman(1983)]{Yellowbook} Lichtenberg, A.~J., \& Lieberman, M.~A.\ 1983, Regular and Chaotic Dynamics (Applied Mathematical Sciences), (New York: Springer)

\bibitem[Lin et al.(1996)]{Lin1996} Lin, D.~N.~C., Bodenheimer, P., \& Richardson, D.~C.\ 1996, \nat, 380, 606 

\bibitem[Lithwick \& Wu(2011)]{LithWu11} Lithwick, Y., \& Wu, Y.\ 2011, \apj, 739, 31 

\bibitem[Malhotra(1990)]{Malhotra1990} Malhotra, R.\ 1990, \icarus, 87, 249 

\bibitem[Marcy et al.(1998)]{Marcy1998} Marcy, G.~W., Butler, R.~P., Vogt, S.~S., Fischer, D., \& Lissauer, J.~J.\ 1998, \apjl, 505, L147 

\bibitem[Marcy et al.(2001)]{Marcy2001} Marcy, G.~W., Butler, R.~P., Fischer, D., et al.\ 2001, \apj, 556, 296 

\bibitem[Mart{\'{\i}} et al.(2013)]{Marti2013} Mart{\'{\i}}, J.~G., Giuppone, C.~A., \& Beaug{\'e}, C.\ 2013, \mnras, 433, 928 

\bibitem[Masset \& Snellgrove(2001)]{MassetSnellgrove2001} Masset, F., \& Snellgrove, M.\ 2001, \mnras, 320, L55 

\bibitem[Mayor \& Queloz(1995)]{MayorQueloz1995} Mayor, M., \& Queloz, D.\ 1995, \nat, 378, 355 

\bibitem[Morbidelli \& Giorgilli(1990a)]{Morby1990a} Morbidelli, A., \& Giorgilli, A.\ 1990a, Celestial Mechanics and Dynamical Astronomy, 47, 145

\bibitem[Morbidelli \& Giorgilli(1990b)]{Morby1990b} Morbidelli, A., \& Giorgilli, A.\ 1990b, Celestial Mechanics and Dynamical Astronomy, 47, 173  

\bibitem[Morbidelli(1993)]{Morby1993} Morbidelli, A.\ 1993, \icarus, 105, 48 

\bibitem[Morbidelli(2002)]{MorbyBook} Morbidelli, A.\ 2002, Modern Celestial Mechanics: Aspects of Solar System Dynamics (London: Taylor \& Francis)

\bibitem[Morbidelli \& Crida(2007)]{MorbyCrida2007} Morbidelli, A., \& Crida, A.\ 2007, \icarus, 191, 158 

\bibitem[Morbidelli et al.(2007)]{Morby2007} Morbidelli, A., Tsiganis, K., Crida, A., Levison, H.~F., \& Gomes, R.\ 2007, \aj, 134, 1790 

\bibitem[Morbidelli et al.(2008)]{Morby2008} Morbidelli, A.,  Levison, H.~F., \& Gomes, R.\ 2008, The Solar System Beyond Neptune, 275 

\bibitem[Morbidelli(2013)]{Morby2013} Morbidelli, A.\ 2013, Planets, Stars and Stellar Systems.~Volume 3: Solar and Stellar Planetary Systems, 63 

\bibitem[Murray(1986)]{Murray1986} Murray, C.~D.\ 1986, \icarus, 65, 70 

\bibitem[Murray \& Dermott(1999)]{MD99} Murray, C.~D., \& Dermott, S.~F.\ 1999, Solar System Dynamics (Cambridge: Cambridge University Press)

\bibitem[Murray et al.(1985)]{Murray1985} Murray, N.~W., Lieberman, M.~A., \& Lichtenberg, A.~J.\ 1985, \pra, 32, 2413 

\bibitem[Murray \& Holman(1997)]{MurrayHolman1997} Murray, N., \& Holman, M.\ 1997, \aj, 114, 1246 

\bibitem[Murray \& Holman(1999)]{MurrayHolman1999} Murray, N., \& Holman, M.\ 1999, Science, 283, 1877 

\bibitem[Murray et al.(2002)]{Murray2002} Murray, N., Paskowitz, M., \& Holman, M.\ 2002, \apj, 565, 608 

\bibitem[Nelson(2005)]{Nelson2005} Nelson, R.~P.\ 2005, \aap, 443, 1067 

\bibitem[Nelson \& Gressel(2010)]{NelsonGressel2010} Nelson, R.~P., \& Gressel, O.\ 2010, \mnras, 409, 639 

\bibitem[Nesvorn{\'y} \& Morbidelli(1998)]{NesvornyMorbidelli1998} Nesvorn{\'y}, D., \& Morbidelli, A.\ 1998, \aj, 116, 3029 

\bibitem[Nesvorn{\'y}(2011)]{Nesvorny2011} Nesvorn{\'y}, D.\ 2011, \apjl, 742, L22 

\bibitem[Nesvorn{\'y} \& Morbidelli(2012)]{NesvornyMorbidelli2012} Nesvorn{\'y}, D., \& Morbidelli, A.\ 2012, \aj, 144, 117 

\bibitem[Ogilvie \& Lubow(2003)]{Ogilvie2003} Ogilvie, G.~I., \& Lubow, S.~H.\ 2003, \apj, 587, 398 

\bibitem[Papaloizou et al.(2007)]{Papaloizou2007} Papaloizou, J.~C.~B., Nelson, R.~P., Kley, W., Masset, F.~S., \& Artymowicz, P.\ 2007, Protostars and Planets V, 655 

\bibitem[Paardekooper \& Papaloizou(2008)]{Paardekooper2008} Paardekooper, S.-J., \& Papaloizou, J.~C.~B.\ 2008, \aap, 485, 877 

\bibitem[Paardekooper \& Papaloizou(2009)]{Paardekooper2009} Paardekooper, S.-J., \& Papaloizou, J.~C.~B.\ 2009, \mnras, 394, 2283 

\bibitem[Peale(1976)]{Peale1976} Peale, S.~J.\ 1976, \araa, 14, 215 

\bibitem[Peale(1986)]{Peale1986} Peale, S.~J.\ 1986, Satellites, 159 (Tucson, AZ: Arizona University Press)

\bibitem[Quinn et al.(1991)]{Quinn1991} Quinn, T.~R., Tremaine, S., \& Duncan, M.\ 1991, \aj, 101, 2287

\bibitem[Rasio \& Ford(1996)]{RasioFord1996} Rasio, F.~A., \& Ford, E.~B.\ 1996, Science, 274, 954 

\bibitem[Raymond et al.(2008)]{Raymond2008} Raymond, S.~N., Barnes, R., \& Gorelick, N.\ 2008, \apj, 689, 478 

\bibitem[Raymond et al.(2009)]{Raymond2009} Raymond, S.~N., Armitage, P.~J., \& Gorelick, N.\ 2009, \apjl, 699, L88 

\bibitem[Rein \& Papaloizou(2009)]{Rein2009} Rein, H., \& Papaloizou, J.~C.~B.\ 2009, \aap, 497, 595 

\bibitem[Rivera \& Lissauer(2001)]{RiveraLissauer2001} Rivera, E.~J., \& Lissauer, J.~J.\ 2001, \apj, 558, 392 

\bibitem[Rivera et al.(2005)]{Rivera2005} Rivera, E.~J., Lissauer, J.~J., Butler, R.~P., et al.\ 2005, \apj, 634, 625 

\bibitem[Rivera et al.(2010)]{Rivera2010} Rivera, E.~J., Laughlin, G., Butler, R.~P., et al.\ 2010, \apj, 719, 890 

\bibitem[Safronov(1969)]{Safronov} Safronov, V.~S.\ 1969, Evolution of the Protoplanetary Cloud and Formation of the Earth and the Planets (Moscow: Nauka Press) Trans. NASA TTF 677, 1972. 

\bibitem[Sessin \& Ferraz-Mello(1984)]{SessinFerrazMello1984} Sessin, W., \& Ferraz-Mello, S.\ 1984, Celestial Mechanics, 32, 307 

\bibitem[Sidlichovsky(1990)]{Sid1990} Sidlichovsky, M.\ 1990, Celestial Mechanics and Dynamical Astronomy, 49, 177

\bibitem[Snellgrove et al.(2001)]{Snellgrove2001} Snellgrove, M.~D., Papaloizou, J.~C.~B., \& Nelson, R.~P.\ 2001, \aap, 374, 1092 

\bibitem[Sussman \& Wisdom(1988)]{SussmanWisdom1988} Sussman, G.~J., \& Wisdom, J.\ 1988, Science, 241, 433 
\bibitem[Sussman \& Wisdom(1992)]{SussmanWisdom1992} Sussman, G.~J., \& Wisdom, J.\ 1992, Science, 257, 56 

\bibitem[Thommes \& Lissauer(2003)]{ThommesLissauer2003} Thommes, E.~W., \& Lissauer, J.~J.\ 2003, \apj, 597, 566 

\bibitem[Timofeev(1978)]{Timofeev1978} Timofeev, A.~V.\ 1978, Fizika Plazmy, 4, 826

\bibitem[Tsiganis et al.(2005)]{Tsiganis2005} Tsiganis, K., Gomes, R., Morbidelli, A., \& Levison, H.~F.\ 2005, \nat, 435, 459 

\bibitem[Veras(2007)]{Veras2007} Veras, D.\ 2007, Celestial Mechanics and Dynamical Astronomy, 99, 197 

\bibitem[Wang \& Uhlenbeck(1945)]{Wang1945} Wang, M.~C., \& Uhlenbeck, G.~E.\ 1945, Reviews of Modern Physics, 17, 323

\bibitem[Ward(1997)]{Ward1997} Ward, W.~R.\ 1997, \icarus, 126, 261 

\bibitem[Wetherill \& Stewart(1989)]{Wetherill} Wetherill, G.~W., \& Stewart, G.~R.\ 1989, \icarus, 77, 330

\bibitem[Wisdom(1983)]{Wisdom1983} Wisdom, J.\ 1983, \icarus, 56, 51 

\bibitem[Wisdom(1985)]{Wisdom1985} Wisdom, J.\ 1985, \icarus, 63, 272 

\bibitem[Wisdom(1986)]{Wisdom1986} Wisdom, J.\ 1986, Celestial Mechanics, 38, 175 

\bibitem[Wright et al.(2011)]{Wright2011} Wright, J.~T., Fakhouri, O., Marcy, G.~W., et al.\ 2011, \pasp, 123, 412 

\bibitem[Yoder(1973)]{Yoder1973} Yoder, C.~F.\ 1973, Ph.D.~Thesis,  

\bibitem[Yoder(1979)]{Yoder1979} Yoder, C.~F.\ 1979, \nat, 279, 767

\bibitem[Zhou \& Sun(2003)]{ZhouSun2003} Zhou, J.-L., \& Sun, Y.-S.\ 2003, \apj, 598, 1290 

\end{thebibliography}
\end{document}